\documentclass[traditabstract]{aa}  
\usepackage[bf]{caption}
\usepackage{graphicx}
\usepackage{subfig}
\usepackage{float}
\usepackage{newclude}
\usepackage{url}
\usepackage{multirow}
\usepackage{enumerate}
\usepackage{txfonts}
\usepackage{natbib}
\bibpunct{(}{)}{;}{a}{}{,} 
\newcommand{\farc}{\hbox{$.\!\!^{\prime\prime}$}} 
\newcommand{\erg}{$\,10^{-17}\mathrm{erg\,s^{-1}\,cm^{-2}}$}
\newcommand{\oii}{[\ion{O}{II}]}
\newcommand{\oiii}{[\ion{O}{III}]}
\newcommand{\nii}{[\ion{N}{II}]}
\newcommand{\ha}{H$\alpha$}
\newcommand{\hb}{H$\beta$}
\newcommand{\msy}{$\rm{M_{\sun}\,yr^{-1}}$}

\begin{document}
    \titlerunning{The solar-metallicity host galaxy of GRB 110918A}
    \title{The low-extinction afterglow in the solar-metallicity host galaxy of $\gamma$-ray burst 110918A \thanks{Based on observations made with telescopes at the European Southern Observatory at La Silla/Paranal, Chile under program 090.A-0760(A).}}
   \author{J.~Elliott\inst{1}
	\and 
	  T.~Kr{\"u}hler\inst{2}
	\and
	  J.~Greiner\inst{1}
	\and
	  S.~Savaglio\inst{1}
	\and
	  F.~Olivares~E.\inst{1}
  \and
	  A.~Rau\inst{1}
	\and
	  A.~de~Ugarte~Postigo\inst{3,2}
	\and
	  R.~S{\'a}nchez-Ram{\'i}rez\inst{3}
	\and               
	  K.~Wiersema\inst{4}
  \and
	  P.~Schady\inst{1}
  \and
	  D.~A.~Kann\inst{5,1}
	\and
	  R.~Filgas\inst{6,1}
  \and
	  M.~Nardini\inst{7}
	\and
    E.~Berger\inst{8}
  \and
    D.~Fox\inst{9}
  \and
	  J.~Gorosabel\inst{3,11,12} 
	\and
	  S.~Klose\inst{5}
  \and
    A.~Levan\inst{10}
  \and
	  A.~Nicuesa Guelbenzu\inst{5}
  \and
	  A.~Rossi\inst{5}
  \and
    S.~Schmidl\inst{5}
	\and
	  V.~Sudilovsky\inst{1}
  \and
	  N.~R.~Tanvir\inst{4}
	\and                  
	  C.~C.~Th{\"o}ne\inst{3}
}
   \institute{
	  Max-Planck-Institut f{\"u}r extraterrestrische Physik, Giessenbachstra{\ss}e 1, 85748, Garching, Germany.\\
      \email{jonnyelliott@mpe.mpg.de}
	  \and 
	    Dark Cosmology Centre, Niels Bohr Institute, University of Copenhagen, Juliane Maries Vej 30, 2100, Copenhagen, Denmark.
	  \and 
   	    Instituto de Astrof{\'i}sica de Andaluc{\'i}a (IAA-CSIC), Glorieta de la Astronom{\'i}a s/n, E-18008, Granada, Spain.
	  \and 
	    The University of Leicester, Department of Physics and Astronomy, University Road, Leicester, LE1 7RH, United Kingdom.
	  \and 
	    Th\"{u}ringer Landessternwarte Tautenburg, Sternwarte 5, 07778, Tautenburg, Germany.
    \and 
	    Institute of Experimental and Applied Physics, Czech Technical University in Prague, Horsk{\'a} 3a/22, 12800, Prague, Czech Republic.
    \and 
      Universit\`a degli studi di Milano-Bicocca, Piazza della Scienza 3, 20126, Milano, Italy.
    \and 
      Harvard-Smithsonian Center for Astrophysics, 60 Garden Street, Cambridge, MA 02138, USA.
    \and 
      Department of Astronomy \& Astrophysics, 525 Davey Laboratory, Pennsylvania State University, University Park, PA 16802, USA.
    \and 
      Department of Physics, University of Warwick, Coventry CV4 7AL, UK.
    \and 
      Unidad Asociada Grupo Ciencia Planetarias UPV/EHU-IAA/CSIC, Departamento de F\'{\i}sica Aplicada I, E.T.S. Ingenier\'{\i}a, Universidad del Pa\'{\i}s Vasco UPV/EHU, Alameda de Urquijo s/n, E-48013 Bilbao, Spain.
    \and 
      Ikerbasque, Basque Foundation for Science, Alameda de Urquijo 36-5, E-48008 Bilbao, Spain.
    }
   \date{Received 20 December 2012 / 
	 Accepted 30 May 2013}

  \abstract{
Galaxies selected through long $\gamma$-ray bursts (GRBs) could be of fundamental importance when mapping the star formation history out to the highest redshifts. Before using them as efficient tools in the early Universe, however, the environmental factors that govern the formation of GRBs need to be understood. Metallicity is theoretically thought to be a fundamental driver in GRB explosions and energetics, but is still, even after more than a decade of extensive studies, not fully understood. This is largely related to two phenomena: a dust-extinction bias, that prevented high-mass and thus likely high-metallicity GRB hosts to be detected in the first place, and a lack of efficient instrumentation, that limited spectroscopic studies including metallicity measurements to the low-redshift end of the GRB host population.
The subject of this work is the very energetic GRB 110918A ($E_{\gamma,\mathrm{iso}}=1.9\times10^{54}\,\mathrm{erg}$), for which we measure a redshift of $z=0.984$. GRB~110918A gave rise to a luminous afterglow with an intrinsic spectral slope of $\beta=0.70$, which probed a sight-line with little extinction ($A^{\mathrm{GRB}}_{V}=0.16\,\mathrm{mag}$) and soft X-ray absorption ($N_{\rm{H,X}} = (1.6 \pm 0.5) \times 10^{21} \rm{cm}^{-2}$) typical of the established distributions of afterglow properties. Photometric and spectroscopic follow-up observations of the galaxy hosting GRB~110918A, including optical/NIR photometry with GROND and spectroscopy with VLT/X-shooter, however, reveal an all but average GRB host in comparison to the $z\sim1$ galaxies selected through similar afterglows to date. It has a large spatial extent with a half-light radius of $R_{\frac{1}{2}} \sim 10$~kpc, the highest stellar mass for $z<1.9$ ($\log(M_*/M_{\sun}) = 10.68\pm0.16$), and an \ha-based star formation rate of $\rm SFR_{H\alpha}=41^{+28}_{-16}\,$\msy. We measure a gas-phase extinction of $A^{\mathrm{gas}}_{V}\sim1.8\,\mathrm{mag}$ through the Balmer decrement and one of the largest host-integrated metallicities ever of around solar using the well-constrained ratios of \nii/\ha, and \nii/\oii~(12 + log(O/H) = $8.93 \pm 0.13$ and $8.85^{+0.14}_{-0.18}$, respectively). This presents one of the very few robust metallicity measurements of GRB hosts at $z \sim 1$, and establishes that GRB hosts at $z \sim1$ can also be very metal rich. It conclusively rules out a metallicity cut-off in GRB host galaxies and argues against an anti-correlation between metallicity and energy release in GRBs.
  }

  \keywords{Gamma-ray burst: general, Gamma-ray burst: individual: GRB 110918A, ISM: general, Galaxies: abundances, Galaxies: photometry, Galaxies: star formation
            }

  \maketitle

\section{Introduction}
\label{sec:Introduction}

During their prompt emission, long gamma-ray bursts (GRBs) are the brightest objects in the Universe, easily reaching isotropic-equivalent luminosities as high as $\sim10^{54}\,\mathrm{erg\, s^{-1}}$. Their observed association to supernovae events~\citep[e.g.,][]{Galama98a,Hjorth03a, Stanek03a,Matheson03a,DellaValle11a,Hjorth12a} has tightly linked them to the death of massive stars. The GRB itself is then believed to result from accretion of matter onto the newly formed rapidly rotating black hole or compact object in the collapsar model~\citep{Woosley93a, Paczynski98a, MacFadyen99a}. The lack of hydrogren and helium in the spectra of GRB-supernovae classify them as type Ic, supporting the notion that GRB progenitors are likely Wolf-Rayet like stars~\citep[for a review of supernova classifications see, e.g.,][]{Filippenko97a}. Given that these type of stars undergo vigorous mass loss from stellar winds, metallicity constraints ($Z<0.3\,\mathrm{Z_{\sun}}$) on the progenitor are postulated to ensure that an accretion disk is still formed around the black hole~\citep{Hirschi05a, Yoon05a, Woosley06a}. 

The possible association of long GRBs with massive stars supported the idea that they could be used as complementary and independent tracers of star formation, especially at high redshifts ($z\gtrsim4$), due to their very high luminosities~\citep[see e.g.,][]{Daigne06a,Li08a,Kistler09a,Ishida11a}. However, to have full confidence in these studies the intrinsic evolutionary effects in long GRB production must be understood and the galactic environments preferred by the progenitor need to be quantified~\citep[e.g.,][]{Butler10a,Wang11a,Salvaterra12a,Robertson12a,Elliott12a}. Of particular interest is the relation between the galaxies selected by GRBs and the star formation weighted population of field galaxies. To be direct and unbiased tracers of star formation, the relative rates of GRBs in galaxies of various physical properties should be the same in galaxies taken from samples that trace the global star formation density at a given redshift. Studies based on these galaxy samples are most commonly performed at $z \lesssim 1.5$, where the star formation of field galaxies is largely recovered by state-of-the-art deep-field surveys.

Initial work showed that many long GRB host galaxies had a low mass, low metallicity, as well as blue colours and were actively star forming~\citep[see e.g.,][]{Fruchter99a, LeFloch03a, Berger03a, Christensen04a, Tanvir04a}. This seemed directly in line with the requirements of the collapsar model. Further work carried out with larger samples~\citep[e.g.][]{Savaglio09a}, showed again similar characteristics. However, at a given mass and star formation rate, long GRB hosts were also found to be no different to the normal population of star forming galaxies at the same redshift \citep{Mannucci11a}. However, these initial studies neglected the contribution from galaxies hosting dust-extinguished afterglows, often termed dark bursts \citep[e.g.,][]{Perley09a, Greiner11a}. Galaxies hosting dark bursts are systematically more massive and have a higher dust-content than the previously-established population localized with optically bright afterglows~\citep{Kruehler11a, Hjorth12a, Rossi12a, Perley13a, Christensen11a, deUgartePostigo12a}.
Despite the inclusion of this more evolved galaxy population, there are still less GRBs in massive galaxies than expected based on their contribution to the overall star formation rate (SFR), at least at $z < 1.5$ \citep{Perley13a}, indicative of a GRB explosion mechanism dependent on metallicity. It is however important to note that these above conclusions are inferred indirectly through stellar mass as a metallicity proxy, and while the photometric samples of GRB hosts have reached integrated number statistics of 100 and above \citep[e.g.,][]{Hjorth12a, Perley13a}, the most crucial measurement of gas-phase metallicity has only been performed in a hand-full cases at $z \gtrsim 1$ \citep{Levesque10d, Kruehler12a}. 

Only a few host galaxies with substantial gas-phase metallicities around or above solar~\citep[e.g.,][]{Levesque10b} that directly violate the proposed cut-off in galaxy metallicity have been observed to date. There is thus still lively debate in the literature about the nature of GRB hosts, and their relation to the star formation weighted galaxy population as a whole \citep[e.g.,][]{Niino11a, Mannucci11a, Kocevski11a, Graham12a}. GRB hosts with high stellar mass and high global metallicity are hence of primary interest for GRB host studies, as they directly probe this allegedly forbidden parameter space. A robust understanding of the galactic environments in which GRBs form would then add confidence in their use as cosmological probes, beyond the limits of deep survey studies \citep[e.g.,][]{Tanvir12a, Basa12a}. 

Here, we present spectroscopy and photometry of the host galaxy and afterglow of the luminous GRB 110918A, detected on the 18th of September 2011 at $T_{0}$=21:26:57 UT \citep{Hurley11a}. This burst had one of the highest fluences of any GRB observed over the last 20 years~\citep[together with GRB 021206;][]{Wigger08a} and had the highest peak flux ever detected by {\it Konus-Wind} \citep[][Frederiks et al. 2013 in prep.]{Golenetskii11a,Frederiks11a}, located at a redshift of $z=0.98$. The massive, metal rich host galaxy and unobscured afterglow of GRB 110918A challenges the current view of the connection between local and global environments and allow us to investigate the preferred conditions for the formation of a long GRB.

The paper is arranged as follows: first we describe the observations carried out by both ground and space based instruments and their corresponding reduction in Sect.~\ref{sec:Observations}. Second, the resulting properties ascertained from the SEDs and spectra of the GRB and its host are described in Sect.~\ref{sec:Results}. Finally, we discuss our findings and their implications for the population of long GRBs in Sect.~\ref{sec:Discussion} and conclude in Sect.~\ref{sec:Conclusion}. We adopt the convention that the GRB flux density is described by $F_{\nu}\left(t\right)\propto t^{-\alpha}\nu^{-\beta}$, reported errors are at the $1\sigma$ confidence level and we assume a $\Lambda$CDM cosmology: $H_{0} = 71\,\mathrm{km\,s^{-1}\,Mpc^{-1}}$, $\Omega_{M} = 0.27$ and $\Omega_{\Lambda}=0.73$. We use a \citet{Chabrier03a} initial mass function and abundances throughout the text.

\section{Observations and Data Reduction}
\label{sec:Observations}

\subsection{Swift -XRT Spectra}
\label{sec:Observations:subsec:Afterglow:subsubsec:XRT}
At the time of the IPN trigger {\it Swift}~\citep{Gehrels04a} was both in the South Atlantic Anomaly and Earth-occulted, and so no trigger~\citep{Krimm11a} was initiated in the Burst Alert Telescope~\citep[][BAT]{Barthelmy05a}. However, the {\it Swift} X-ray telescope~\citep[][XRT]{Burrows05a} began observing the field of GRB 110918A (see Fig.~\ref{fig:fits_grb}) at $T_{0}+107.4\,\mathrm{ks}$ until $\sim40\,\mathrm{d}$ later. The XRT spectrum shows no signs of spectral evolution, remaining with a constant hardness ratio of $\sim0.85$ for its entire emission. We extract a spectrum at the time interval of $T_{0}+140\, \rm ks$ to $T_{0}+250\, \rm ks$ to coincide with our optical/NIR wavelength observations (see Fig.~\ref{fig:sed_grb_broadband}). The XRT spectral data were obtained from the public {\it Swift} archive and were regrouped to ensure at least 20 counts per bin in the standard manner, using the \emph{grappha} task from the HEAsoft package with response matrices from CALDB \texttt{v20120209}. We assume a Galactic hydrogen column of $\mathrm{N^{Gal}_{H,X}}=1.68\times10^{20}\,\mathrm{cm^{-2}}$~\citep{Kalberla05a} in the direction of the burst.

\subsection{GROND Optical/NIR Photometry}
\label{sec:Observations:subsec:Afterglow:subsubsec:GROND}

\begin{figure}
 \includegraphics[trim=0 0 0 0, clip=True, width=9.0cm]{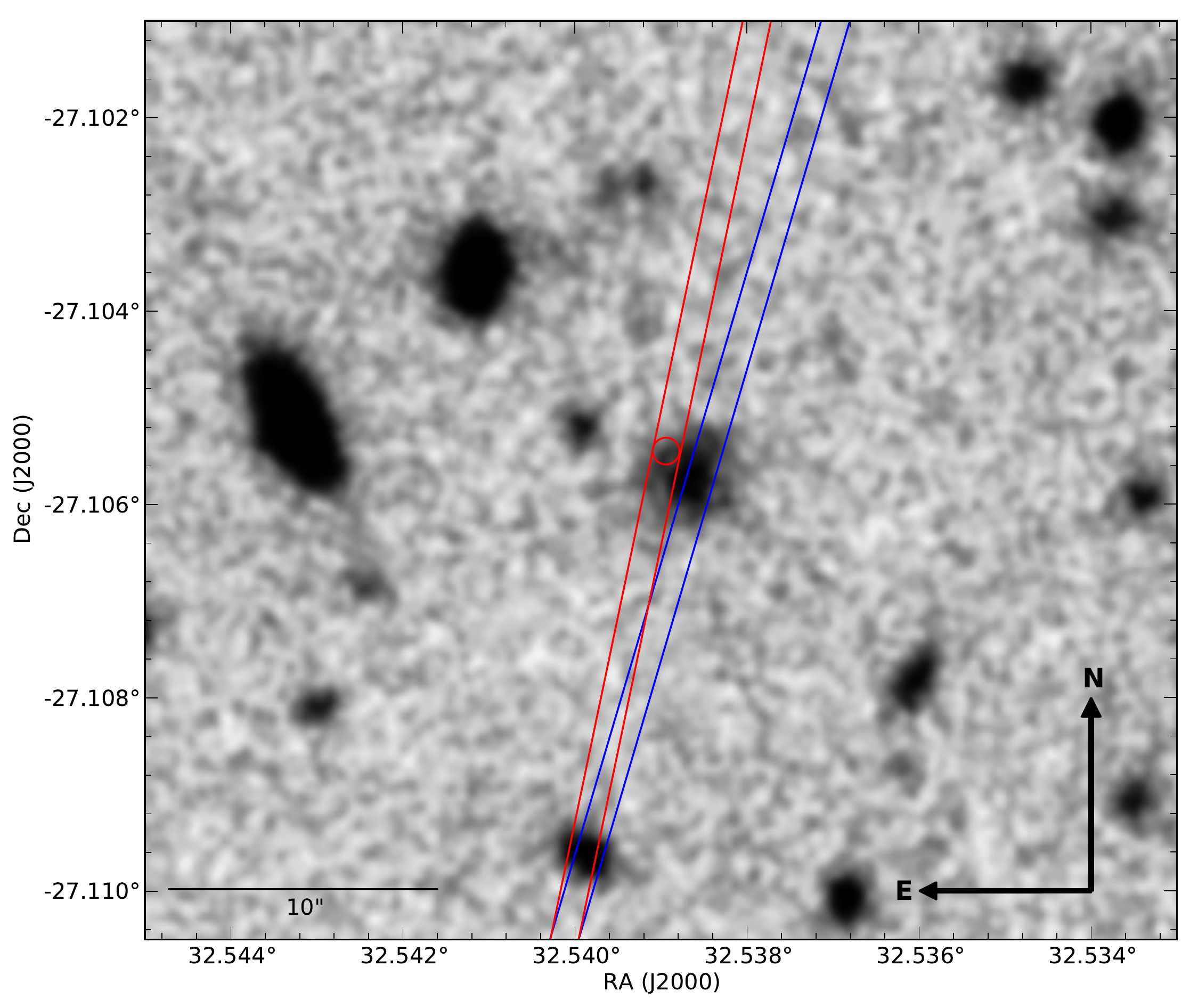}
 \caption{A GROND 60 minute stacked $i'$ band image of the GRB 110918A host galaxy. The slit arrangement used in the acquisition image is shown for both the OSIRIS afterglow spectrum (red line) and host spectrum (blue line).}
 \label{fig:fits_grb}
\end{figure}

The Gamma-Ray Burst Optical Near-infrared Detector~\citep[GROND;][]{Greiner08a} mounted at the MPG/ESO $2.2~\mathrm{m}$ telescope at La Silla, Chile, began its follow-up campaign of GRB 110918A $29.2~\mathrm{hrs}$ after the trigger simultaneously in the $g^{\prime}r^{\prime}i^{\prime}z^{\prime}JHK_{s}$ filters \citep{Elliott11a}. A mosaic of 5 pointings was carried out to cover the full IPN error box ($\sim20\arcmin \mathrm\times20\arcmin$) and the GRB optical afterglow candidate was detected at the location R.A. (J2000) = $02^{h}10^{m}09.34^{s}$, Dec. (J2000) = $-27\degr06\arcmin19.7\arcsec$, in GROND's NIR chips, located just outside the IPN error box, consistent with the X-ray \citep{Mangano11a} and optical \citep{Tanvir11a} position with an uncertainty of $0.2\arcsec$. GROND observations continued for over 1 month after the GRB trigger and an underlying host was discovered \citep[see also][]{Oksanen11b}. Deep images of 3600 s in the NIR and 4500 s in the optical were obtained with GROND of the host galaxy at $T_{0}+36.37\,\mathrm{d}$.   

Image reduction and photometry of the GROND observations were carried out using standard IRAF tasks~\citep{Tody93a} in the way outlined in~\citet{Kruehler08a} and~\citet{Yoldas08a}. In brief, a point-spread function (PSF) was obtained from the bright stars within the field and applied to the afterglow photometry. The absolute calibration of the optical photometry was achieved by observing a Sloan Digital Sky Survey (SDSS) field~\citep{2011ApJS..193...29A} at R.A. (J2000) = $01^{h}40^{m}00.0^{s}$, Dec. (J2000) = $-18\degr03\arcmin00.0\arcsec$ and the GRB field consecutively. The NIR absolute calibration was obtained from the Two Micron Sky Survey (2MASS) stars~\citep{Skrutskie06a} within the field of the GRB. As a result of the extension of the galaxy (see Fig.~\ref{fig:fits_grb}) an aperture size of $3.2\arcsec$ was used on the deep host galaxy images, within which the aperture flux flattened in a curve-of-growth analysis and the zero points were correspondingly corrected.

The scope of this paper does not involve a full analysis of the afterglow emission. However, afterglow flux measurements are required at certain time intervals to implement slit loss corrections to the optical spectra and to determine the local extinction of the afterglow. Therefore, we only present the required data and direct the reader to future work for a full analysis of the afterglow\footnote{For the interested reader the afterglow light curves are found in the Appendix.}. For the slit loss correction of the afterglow, we obtain the following brightnesses in the AB system at a mid time of $T_{0}+2.2\,\mathrm{d}$: $g'=20.50\pm0.05\,\mathrm{mag}$, $r'=20.20\pm0.04\,\mathrm{mag}$, $i'=19.96\pm0.04\,\mathrm{mag}$, $z'=19.83\pm0.07\,\mathrm{mag}$, $J=19.46\pm0.09\,\mathrm{mag}$, $H=19.04\pm0.10\,\mathrm{mag}$, $K_{s}=18.66\pm0.19\,\mathrm{mag}$. The magnitudes are uncorrected for a Galactic dust reddening of $E(B-V)^{Gal}=0.020\,\mathrm{mag}$ corresponding to an extinction of $A^{\mathrm{Gal}}_{V}=0.062\,\mathrm{mag}$ for $R_{V}=3.1$~\citep{Schlegel98a}. 

\subsection{WFI Optical Photometry}
\label{sec:Observations:subsec:Afterglow:subsubsec:WFI}

Further deep observations of the host galaxy were made 392 d after the trigger with the Wide Field Imager~\citep[][WFI]{Baade99a}, also mounted on the MPG/ESO 2.2m telescope, in the standard broadband filters $\mathrm{BB}\#B\mathrm{/123\_ESO878}$ ($B$) and $\mathrm{BB}\#U\mathrm{/150\_ESO878}$ ($U$). Two sets of images were taken, the first on 25 October 2012 consisting of 1800 s in $U$ and 600 s in $B$ and the second set was on 26 October 2012, consisting of 150 s in $U$ and 75 s in $B$. A calibration field was obtained on 26 October 2012 in both the $U$ and $B$ filters and the standard field SA113+158\footnote{\url{www.eso.org/sci/observing/tools/standards/Landolt.html}} was used as a primary calibrator. The photometry was carried out in the same way as the GROND images, and the magnitudes converted into the AB system using the ESO magnitude converter\footnote{\url{http://archive.eso.org/mag2flux}}.

\subsection{WISE IR Photometry}
\label{sec:Observations:subsec:Afterglow:subsubsec:WISE}

The Wide-field Infrared Survey Explorer~\citep[][WISE]{Wright10a} All-Sky Source Catalogue\footnote{\url{http://irsa.ipac.caltech.edu}} reveals a source at the position of the host galaxy of GRB 110918A, with an $11\sigma$ and $3\sigma$ detection in the $W1$ and $W2$ bands, centred at 3.4$\mu$m and 4.6$\mu$m respectively.  The \emph{wmpro} magnitudes were used, which are the magnitudes retrieved from profile-fitting photometry (or the magnitude of the 95\% confidence brightness) and converted into the AB system using the WISE conversion factors\footnote{\url{http://wise2.ipac.caltech.edu/docs/release/allsky/expsup/sec4_4h.html}}. Galactic reddening corrections were made using the $A^{\mathrm{Gal}}_{V}$ conversions determined by~\citet{Jarrett12a}.

\subsection{GMOS Optical Spectroscopy}
\label{sec:Observations:subsec:Afterglow:subsubsec:Gmos}
The first spectrum of the afterglow was taken with the Gemini Multi-Object Spectrographs~\citep[][GMOS]{Hook04a} on the Gemini North telescope (Mauna Kea) starting at 12:52 UT on 20 September 2011, 1.6 d after the GRB trigger. Four exposures of 500 s each were obtained using the R400 grism and a 1\arcsec ($\sim8.0\,\mathrm{kpc}$ projected at $z=0.984$) slit width.  Two of the spectra were obtained with a central wavelength of $6000\,$\AA~and the other two with $6050\,$\AA~to cover the detector gaps. In addition, a spatial dither was used to cover the amplifier boundaries. The resulting spectrum covers the range  $3930 - 8170\,$\AA.  We reduced the data with tasks within the {\it Gemini.GMOS} package and {\it IRAF}, \texttt{v1.11}, using flat field and arc lamp frames taken directly before and after the science data.

\subsection{OSIRIS Optical Spectroscopy}
\label{sec:Observations:subsec:HostGalaxy:subsubsec:Osiris}
The second spectrum of the afterglow was obtained using the Optical System for Imaging and low Resolution Integrated Spectroscopy~\citep[][OSIRIS]{Cepa00a} mounted on the $10.4\,\mathrm{m}$ Gran Telescopio Canarias (Roque de los Muchachos) starting at 13:00 UT on 21 September 2011, 2.2 d after the GRB trigger. Three exposures of 900 seconds each were taken using the R500B grism and a 1\arcsec slit width obtained at the parallactic angle. The resulting spectrum covers the range $4400 - 8700\,$\AA. Data were reduced and calibrated using standard procedures in IRAF. The spectrum was flux calibrated using G157-34 as a standard star and the 1D spectrum was scaled to the GROND afterglow photometry to correct for slit losses (correction factor of $\sim1.9$).

A spectrum of the host was obtained on 11 November 2011, $T_{0}+54.1$ d after the GRB trigger. A sequence of three 1200 s exposures were obtained with OSIRIS using the R1000R grism and a $1.0\arcsec$~slit width, covering the wavelength range of $5100\,$\AA~to $10000\,$\AA. The spectrum was flux calibrated using the standard star G191-B2B and corrected for slit losses by scaling the 1D spectrum to the photometry of the host galaxy obtained using GROND (correction factor of $\sim3.5$). 

\subsection{X-Shooter Optical/NIR Spectroscopy}
\label{sec:Observations:subsec:HostGalaxy:subsubsec:Xshooter}
We further observed the host of GRB~110918A with the cross-dispersed echelle spectrograph X-shooter \citep{2011arXiv1110.1944V} on the Very Large Telescope Kueyen (UT2). X-shooter has three individual arms taking spectra simultaneously in the range of $3000\,\mathrm{\AA}$ to $5600\,\mathrm{\AA}$ (UVB arm), $5600\,\mathrm{\AA}$ to $10\,200\,\mathrm{\AA}$ (VIS arm), and $10\,200\,\mathrm{\AA}$ to $24\,800\,\mathrm{\AA}$ (NIR arm). Three different sets of observations were carried out on 17 December 2012, 07 January 2013, and 16 January 2013, respectively, with a position angle of $59^{\circ}$ East of North. They consisted of a pair of nodded frames with exposure times of $1200$~s in each of the UVB/VIS arm and $2\times600$~s in the NIR arm. The slit width was 1\farc{6}, 1\farc{5} and 0\farc{9} yielding a resolution measured on arc-lamp frames of $R \sim \lambda/\Delta\lambda=3200$, 4900, and 5300 in the UVB, VIS and NIR arm, respectively. The NIR slit includes a blocking filter for the $K$-band limiting our effective wavelength coverage to $<20\,500\,\mathrm{\AA}$, but providing lower background levels in the $J$ and $H$-band.

Each of the individual observations were reduced and wavelength- and flux-calibrated separately using standard procedures within the X-shooter pipeline \texttt{v2.0.0} \citep{2006SPIE.6269E..80G} supplied by ESO. The individual two-dimensional frames were then stacked using variance weighting in a heliocentric reference frame, and the one-dimensional spectra were extracted using an optimal extraction method. Given the extent of the target, slit-losses are substantial. Similar to the OSIRIS and GMOS spectroscopy, we scaled the well-detected continuum of the X-shooter data to the available photometric host SED. The consistency between matching factors derived from different photometric data in the individual arms\footnote{We derive factors for the $r'$-, $i'$- and $z'$ -band data in the VIS arm of $3.1 \pm 0.3$, $2.8 \pm 0.3$, $2.8 \pm 0.3$, and $3.5 \pm 0.4$, $3.3 \pm 0.3$ for the $J$ and $H$-band in the NIR arm. The offset between the VIS and NIR arm is readily explained by the smaller slit width in the NIR arm.}, provides confidence that the absolute flux-calibration in the final X-shooter spectrum is accurate to better than $\sim 20$\% over the full wavelength range of interest.
 

\section{Results}
\label{sec:Results}

\subsection{The Afterglow Sight-Line: Dust, Star Formation Rate and Gas}
\label{sec:Results:subsec:broadbandsed}
\label{sec:Results:subsec:afterglowspectra}
\begin{figure}
 \center
  \includegraphics[trim=0cm 0cm 0cm 0cm, clip=True, width=9cm]{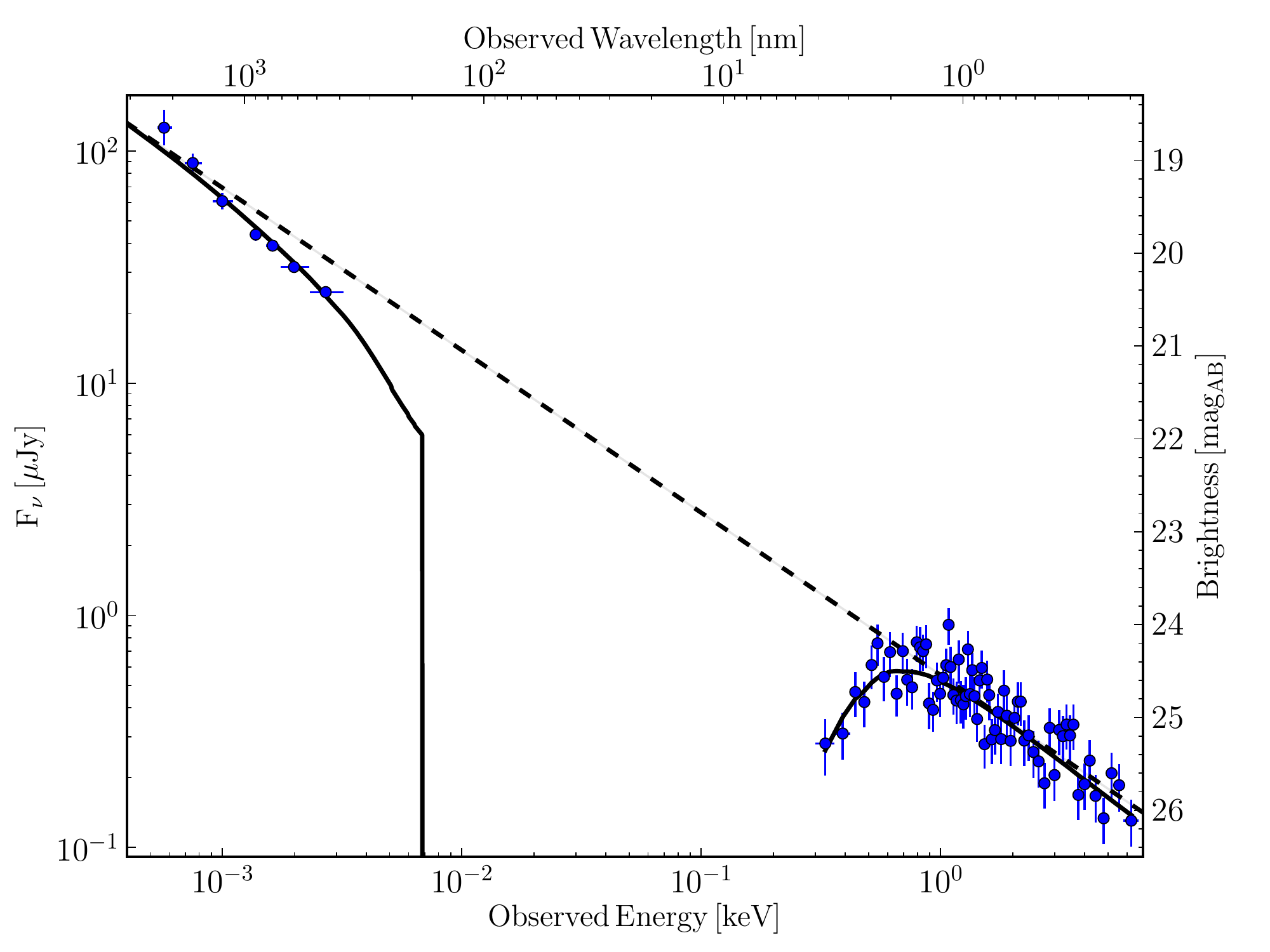}
  \caption{Broadband SED of GRB 110918A, including optical, NIR and X-ray data. The SED was constructed using GROND data at a mid time of $T_{0}+194\,\mathrm{ks}$ and X-ray data between $T_{0}+140\,\rm ks$ and $T_{0}+250\,\rm ks$. The best-fit parameters for a power-law ($\chi^{2}/\mathrm{d.o.f.}=85/73$) are: a spectral slope of $\beta=0.70\pm0.02$, a hydrogen column density of $\rm N^{GRB}_{H,X}=1.56^{+0.52}_{-0.46}\times10^{21}\,cm^{-2}$ and a line of sight extinction of $A^{\mathrm{GRB}}_V=0.16\pm0.06\,\mathrm{mag}$, assuming SMC like dust.}
  \label{fig:sed_grb_broadband}
\end{figure}

A broadband SED was constructed from the optical/NIR GROND photometry (see Sect.~\ref{sec:Observations:subsec:Afterglow:subsubsec:GROND}) at a mid time of $T_{0}+194\,\mathrm{ks}$ and X-ray data between $T_{0}+140\,\rm ks$ and $T_{0}+250\,\rm ks$. The SED was fit in a standard manner \citep[e.g.,][]{Filgas11a}, assuming that the afterglow emission is well described by the standard synchrotron mechanism. 
The best fit single power-law ($\chi^{2}/\mathrm{d.o.f.}=85/73$) is shown in Fig.~\ref{fig:sed_grb_broadband} with a spectral slope of $\beta=0.70\pm0.02$, a hydrogen column density of $\rm N^{GRB}_{H,X}=1.56^{+0.52}_{-0.46}\times10^{21}\,cm^{-2}$ and a line of sight extinction of $A^{\mathrm{GRB}}_V=0.16\pm0.06\,\mathrm{mag}$, assuming a Small Magellanic Cloud (SMC) like dust with $R_{V}=2.93$ in the parametrization of \citet{1992ApJ...395..130P}. Despite yielding an improved fit, a broken power-law model is not warranted as the number of free parameters increases ($\chi^{2}/\mathrm{d.o.f.}=83/71$), implying the improvement is not statistically significant. Nevertheless, the resulting parameters of the best fit broken power-law, as well as from different dust models, are consistent with the uncertainties of the power-law values below the break at $\sim 0.6$~keV and do not alter our conclusions.

Using the procedure outlined by~\citet{deUgartePostigo12a} with the two afterglow spectra obtained with GMOS ($T_{0}+1.6\mathrm{d}$) and OSIRIS ($T_{0}+2.2\mathrm{d}$), we detect the transition of several metal ions including \ion{Fe}{II}, \ion{Mg}{II} and \ion{Mg}{I} at a common redshift of $z=0.984\pm0.001$ (see Tables \ref{tab:phot_par} and \ref{tab:spectra_equivalentwidths}), consistent with galactic winds or star bursting periods~\citep{Fynbo09a, Nestor11a, Christensen11a, RodriguezHidalgo12a}. In comparison with a long GRB sample~\citep{deUgartePostigo12a} we find that it has stronger absorption features than 80\% of the sample.

\subsection{The Host's Stellar Component: Dust Attenuation, Star Formation Rate and Stellar Mass}
\label{sec:Results:subsec:HostSED:subsubsec:PhotometricSED}

\begin{figure}
 \centering
 \includegraphics[trim=0.5cm 0cm 0cm 0cm, clip=True, width=9.0cm]{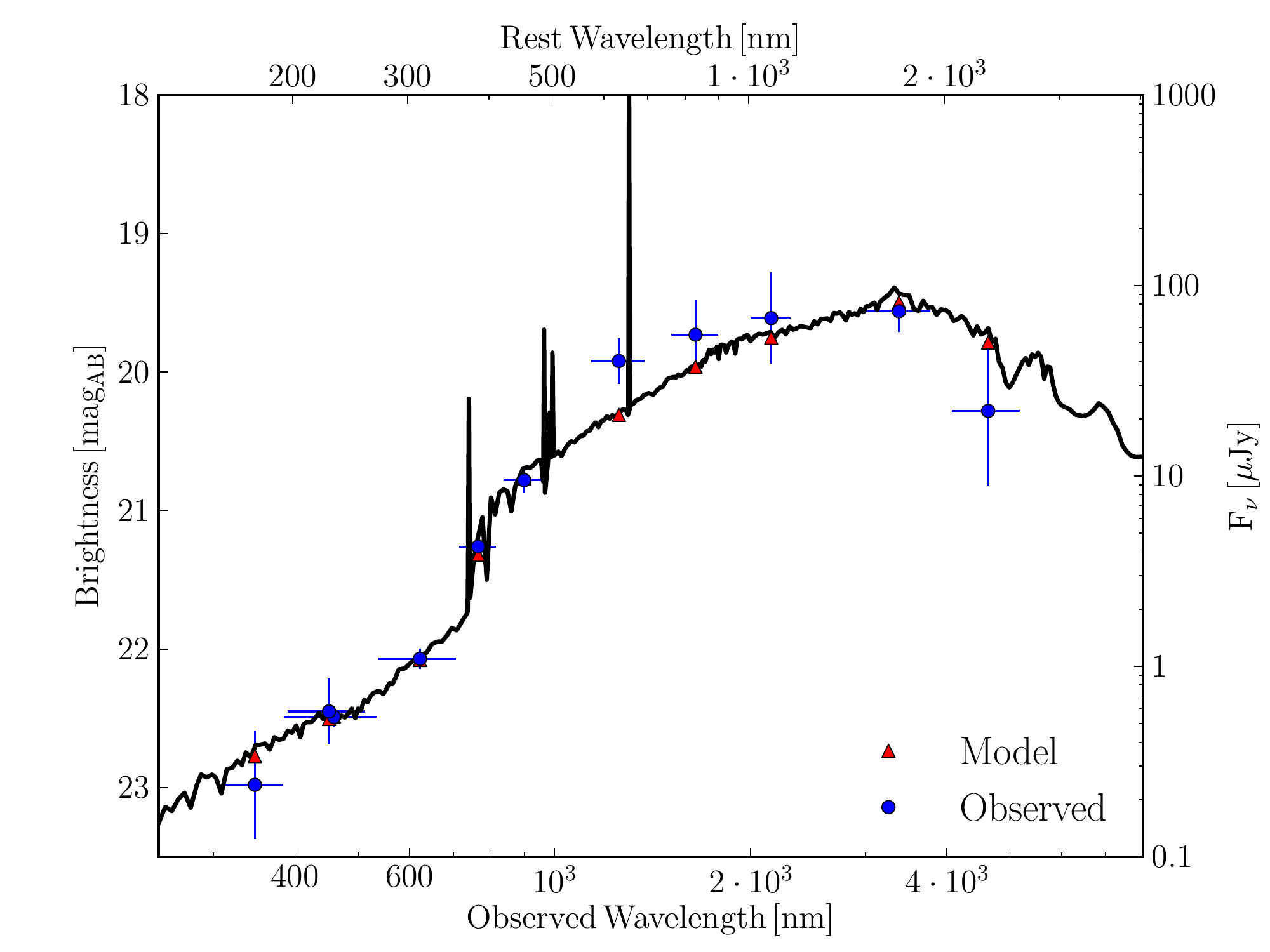}
 \caption{The SED of the host of GRB 110918A obtained using GROND, WFI and WISE data, amounting to 11 filters: $UBg'r'i'z'JHK_{s}W1W2$ from left to right. The best-fit spectrum is depicted in black.}
 \label{fig:_sedhost}
\end{figure}

The host of GRB 110918A was detected in 11 different filters ranging from the ultra-violet to $4.5\mu$m, yielding a well-sampled photometric SED (see Fig.~\ref{fig:_sedhost} and Table~\ref{tab:phot_host}). To estimate the global properties of the host galaxy we employed standard techniques that use stellar population synthesis to estimate stellar masses, as outlined thoroughly in~\citet{Ilbert09a}. We constructed a grid of galaxy templates based on the models taken from~\citet{Bruzual03} over a wide parameter space consisting of: a range of ages ($0-1.35\times10^{9}\mathrm{yr}$), star formation histories ($\propto e^{\tau},\tau=0.1,0.3,1,2,3,5,10,15,30$), reddening values ($E(B-V)=0-0.4\,\mathrm{mag}$), a single attenuation law~\citep[starburst;][]{Calzetti00a} and metallicities ($Z=0.004,0.008,0.02$). Emission lines were also included, whereby the emission lines were estimated from the predicted UV luminosity and converted to a star formation rate using \citet{Kennicutt98a}. For each template a SED was constructed for the filters required and a $\chi^{2}$ was calculated using the Photometric Analysis for Redshift Estimate routines, LePHARE~\footnote{\url{www.cfht.hawaii.edu/~arnouts/LEPHARE/lephare.html}} \texttt{v2.2}~\citep{Arnouts99a, Ilbert06a}. The best fit template was the one that gave the minimum $\chi^{2}$ and the corresponding uncertainties for each parameter were obtained from the grid of $\chi^{2}$ values. Systematic uncertainties of up to an average of $0.2-0.3\,\mathrm{dex}$ are expected in the stellar mass value due to the adopted stellar population models and extinction laws~\citep[see e.g.,][and references therein]{Kruehler11a}.
The filter response curves for the $W1$ and $W2$ bands were obtained from~\citet{Wright10a} and for the $U$ and $B$ bands from the ESO web pages\footnote{\url{www.eso.org/sci/facilities/lasilla/instruments/wfi/inst/filters}}.
The results of the best fit template, which had a $\chi^{2}/\mathrm{\#\,Bands}=5.4/11$, had the following parameters: a mass of $\log_{10}\left(\mathrm{\frac{M_{*}}{M_{\sun}}}\right)=10.68\pm0.16$, a star formation rate $\mathrm{SFR_{SED}}=66^{+50}_{-30}\,\mathrm{M_{\odot}}$, a reddening of $E(B-V)^{\rm{stars}}=0.26\pm0.15\,\mathrm{mag}$, and a starburst age of $\tau=0.7^{+1.4}_{-0.4}\,\mathrm{Gyr}$.

\begin{table}
\caption{Host galaxy magnitudes.}
\begin{center}

  \begin{tabular}{c c c c}
  Filter & Instrument & Magnitude & Uncertainty \\
    &  & $(\mathrm{mag_{AB}})$ & $(\mathrm{mag_{AB}})$ \\
     \hline\hline
    $U$ & WFI & 22.98 & 0.25 \\
    $B$ & WFI & 22.45 & 0.16 \\
    $g'$ & GROND & 22.49 & 0.15 \\
    $r'$ & GROND & 22.07 & 0.05 \\
    $i'$ & GROND & 21.26 & 0.06 \\
    $z$ & GROND & 20.78 & 0.06 \\
    $J$ & GROND & 19.92 & 0.11 \\
    $H$ & GROND & 19.73 & 0.17 \\
    $K_s$ & GROND & 19.61 & 0.22 \\
    $W1$ & WISE & 19.56 & 0.10 \\
    $W2$ & WISE & 20.28 & 0.36 \\
 \hline\hline
  \end{tabular}
 \tablefoot{Corrected for Galactic foreground reddening. The observations in $g'r'i'z'JHK_{s}$ were obtained $36.37\,\mathrm{d}$ after the burst. The $UB$ observations were obtained 392 d after the burst. The $W1$ and $W2$ photometry were obtained prior to the burst by the WISE Survey.}
\label{tab:phot_host}
\end{center}
\end{table}

\subsection{The Host's Gas-Phase Component: Dust Extinction, Star Formation Rate and Metallicity}
\label{sec:Results:subsec:Host:subsubsec:OpticalSpectrum}

The first host spectrum of GRB 110918A was obtained with OSIRIS/GTC in the optical wavelength range, $\sim50$ days after the trigger, and the one second was obtained with X-shooter/VLT more than $460$ days post trigger. The X-shooter spectrum extends our spectral coverage to the NIR and thus to the wavelength range where important tracers of star formation rate and metallicity are located. In summary, we clearly detect the H$\alpha$ and H$\beta$ transition from the Balmer series, as well as the forbidden transitions of \oii($\lambda\lambda$3726, 3729) and \nii($\lambda$6584). The emission lines corresponding to \oiii($\lambda\lambda$4959,5007) are cosmologically redshifted to regions of low sensitivity for both OSIRIS and X-shooter, and are thus not detected.

The velocity profile of the emission lines is clearly resolved by our X-shooter data and spans approximately $500\,\rm{km\,s^{-1}}$ in velocity space (see Fig.~\ref{fig:lines}). It displays a conspicuous two-humped profile, with the two peaks of the emission lines separated by 200~$\rm{km\,s^{-1}}$. However, we do not observe a spatial tilt in the line-shape as would have been expected from a largely rotationally-supported galaxy (unless it is face on), and both peaks appear at the same spatial position in the two-dimensional spectrum. Line fluxes were measured by numerically integrating the available data, and cross-checked with fitting Gaussians. Both procedures return consistent values, and from the X-shooter spectrum, we measure global\footnote{Here and in the following, the error includes both the statistical error of the measurement as well as the error in slit-loss correction.}  emission-lines fluxes of $f_{\oii} = (19.0 \pm 3.1)\times$\erg,  $f_{\rm{H}\beta} = (9.5 \pm 1.9)\times$\erg, $f_{\rm{H}\alpha}= (47.8 \pm 4.9)\times$\erg~and $f_{\ion{[N}{II]}}= (15.3 \pm 3.3)\times$\erg. The OSIRIS spectrum yields $f_{[\ion{O}{II]}]} = (20.0 \pm 2.8)\times$\erg, fully consistent with the X-shooter value.

Assuming case B recombination~\citep{Osterbrock89a} and using the standard values for electron density ($10^{2}\,\mathrm{cm^{-3}} \lesssim n_{\rm e} \lesssim 10^{4}\,\mathrm{cm^{-3}}$) and temperature ($T_{\rm e}\sim10^{4}\,\rm K$), the Balmer ratio of \ha/\hb~implies an $E(B-V)^{\rm{gas}}=0.57^{+0.24}_{-0.22}$ or visual extinction $A_V^{\rm{gas}}=1.8^{+0.8}_{-0.7}$~mag towards the star forming regions assuming a Milky-Way like extinction law\footnote{Different local extinction laws yield comparable results, as there is little difference in the wavelength range of the \ha~and \hb~Balmer lines.}. It is worth mentioning, that this is the luminosity-weighted reddening/extinction of the gas-phase. This value is typically found to be a factor of around two larger than the stellar $E(B-V)^{\rm{stars}}$ from the photometric SED model \citep[e.g.,][]{Calzetti00a}, consistent with our measurements for GRB~110918A. The \ha~line flux implies a SFR of $\rm SFR_{H\alpha}=41^{+28}_{-16}\,$\msy, following \citet[][]{Kennicutt98a} with a \citet{Chabrier03a} IMF.

Using the different emission line ratios, we can measure the gas-phase metallicity of the galaxy hosting GRB~110918A \citep[see e.g.,][ for an extensive summary on those techniques]{2008ApJ...681.1183K}. 
At $z\sim1$ we
we are limited to the diagnostic ratios based on \ha, \oii~and \nii. When using the ratio of \nii~and \ha~as a metallicity tracer, uncertainties in the reddening or chromatic slit-losses are not going to affect the overall results, because the respective lines are located very close in wavelength space. However, \nii/\ha~saturates at high metallicities (\nii/\ha $\sim 0.3$), while \nii/\oii~does not. The final uncertainties are thus comparable in both indicators, and we measure 12 + log(O/H)$_{\rm{N2H\alpha}} = 8.93 \pm 0.13$ and 12 + log(O/H)$_{\rm{N2O2}} = 8.85_{-0.18}^{+0.14}$ using the formulation of \citet{2006A&A...459...85N}. Different calibrations of the strong-line diagnostics yield 12 + log(O/H)$_{\rm{N2H\alpha}} = 8.63\pm0.08$ \citep{2004MNRAS.348L..59P} or 12 + log(O/H)$_{\rm{N2O2}} = 8.86_{-0.14}^{+0.10}$ \citep{2002ApJS..142...35K}. 
The inherent systematic uncertainty 
of typically 0.1-0.2 dex \citep[e.g.,][]{2006A&A...459...85N, 2008ApJ...681.1183K} has not been included. These measurement imply metallicities between 0.9 and 1.7 times solar. All physical parameters of the galaxy hosting GRB~110918A are summarized in Table~\ref{tab:phot_par}.

\begin{table}
\caption{Physical parameters of the galaxy hosting GRB~110918A}
\begin{center}
\begin{tabular}{c c c }
\hline
Quantity & Unit/Method & Value \\
\hline
\hline
 $E(B-V)^{\rm{GRB}}$ & mag & $0.05\pm0.02$ \\
 $N_{\rm{X,H}}^{\rm{GRB}}$ & $10^{21}\,\rm{cm}^{-2}$ & $1.56^{+0.52}_{-0.46}$ \\
EW$_{\rm rest}^{\rm{GRB}}$ & (\AA) \ion{Mg}{II}(2796, 2803) & 6.0 \\
 $\rm SFR^{GRB}_{[OII]}$ & \msy & $2.3\pm0.7$ \\
\hline
Stellar Mass & $\log(M_{*}/M_{\sun})$  & $10.69\pm0.16$ \\ 
Half-light radius & kpc  & $10$ \\ 
 $E(B-V)^{\rm{stars}}$ & mag  &$0.26\pm0.15$ \\
 $\rm SFR_{SED}$ & \msy &  $66^{+50}_{-30}$ \\
$\tau$ & Gyr &  $0.7^{+1.4}_{-0.4}$ \\
\hline
 $SFR_{H\alpha}$ & \msy & $41^{+28}_{-16}$ \\
 $E(B-V)^{\rm{gas}}$ & mag &   $0.57^{+0.24}_{-0.22}$ \\
 12+$\log(\rm{O/H})$ &  \nii/\ha$^{(a)}$  &  $8.93 \pm 0.13$ \\
 12+$\log(\rm{O/H})$ & \nii/\ha$^{(b)}$  &  $8.63 \pm 0.08$ \\
 12+$\log(\rm{O/H})$ &  \nii/\oii$^{(a)}$  &  $8.86^{+0.10}_{-0.14}$ \\
 12+$\log(\rm{O/H})$ &  \nii/\oii$^{(c)}$  &  $8.85^{+0.14}_{-0.18}$ \\
 \hline
  \end{tabular}
 \tablefoot{All values use a Chabrier IMF, and take into account the statistical uncertainty of the measurements, as well as the uncertainty in the slit-loss and dust-correction factor if applicable.
 $^{(a)}$ Following \citet{2006A&A...459...85N}.
 $^{(b)}$ Following \citet{2004MNRAS.348L..59P}.
 $^{(c)}$ Following \citet{2002ApJS..142...35K}. 
}
\label{tab:phot_par}
\end{center}
\end{table}

\section{Discussion}
\label{sec:Discussion}

\subsection{Host Galaxy Identification}
\label{sec:Discussion:subsec:hostidentification}

We have used
absorption lines from metal ions in the afterglow spectrum \citep[see also][for an extensive sample]{Fynbo09a}, and forbidden/recombination lines from the host galaxy to determine the redshift of the GRB \citep[see also][]{Kruehler12b}, but it is in principle possible that the GRB lies at a higher redshift.
To investigate if the host galaxy of GRB 110918A has been misidentified we calculate the commonly used p-value, $p\left(m\right)=1-\exp\left(-\pi r_{i}^{2}\sigma\left(\le m \right) \right)$, which is the probability of finding a galaxy of magnitude $m$ (or brighter) overlapping the GRB within an effective radius $r_{\rm{i}}$, assuming that galaxies are Poisson distributed throughout the sky~\citep{Bloom02a}. This neglects any type of galaxy clustering, however, recent work indicates that GRB locations do not preferentially lie in areas of strong galaxy overabundances~\citep[][]{Cucchiara12a,Sudilovsky13a}. The number of galaxies brighter than $m$ per square arcsecond is given by $\sigma$, taken from~\citet{Bloom02a} and calculated from the work of~\citet{Hogg97a}. 

The burst location of GRB 110918A is seen to be offset from the bright centroid of the host by $12\,\mathrm{kpc}$, however, in comparison to the half-light radius of the host galaxy, $R_{\frac{1}{2}}=10.6\,\mathrm{kpc}$, the offset is consistent with the long GRB population, which has a median offset of $R_{\mathrm{offset}}/R_{\frac{1}{2}}\sim1$~\citep{Bloom02a}. We follow~\citet{Bloom02a} and set $r_{i}=2\times R_{\frac{1}{2}}=2.66\arcsec$. Synthetic $R_{C}$-band \citep{Bessel79a} photometry of the GRB 110918A host galaxy using the best fit galaxy template taken from Sect.~\ref{sec:Results:subsec:HostSED:subsubsec:PhotometricSED}, and the conversions given in~\citet{Rossi12a}, results in $R_{C}=21.7\,\mathrm{mag_{Vega}}$. This yields a probability of chance association of $p=0.01$, making this galaxy highly likely the host of GRB 110918A.


The non-detection of the Lyman forest above $\sim$4500 {\AA} implies a strong upper limit of $z<2.7$. Therefore,
using our knowledge of the strength of spectral features in GRB environments and their distribution ~\citep{deUgartePostigo12a}, we can estimate the likelihood of the GRB having occurred between redshift 1.0 and 2.7 and yet not having detectable absorption lines at the redshift of the host in its spectrum.
We calculate the detection limits for \ion{Mg}{ii} and \ion{C}{iv} doublets as described by~\citet{deUgartePostigo12a} and find that the lines would have to be weaker than
99.7\% of a normal long GRB sample
to have happened at a redshift between 1 and 2.7.
Furthermore, the properties of the absorber (strong \ion{Mg}{II} absorption and vigorous star formation, see Sect.~\ref{sec:Results:subsec:afterglowspectra}), are very common in other afterglow observations, and do not indicate a different physical nature. Combining the arguments presented above, we consider the redshift of the GRB, and accordingly the physical association between GRB and galaxy, robust.


\subsection{Host Environment in the Context of the GRB-Host Population}
\label{sec:Discussion:subsec:HostGRBEnvironment}

\begin{figure}     
 \includegraphics[trim=0cm 0cm 0cm 0cm, clip=True, width=9.5cm]{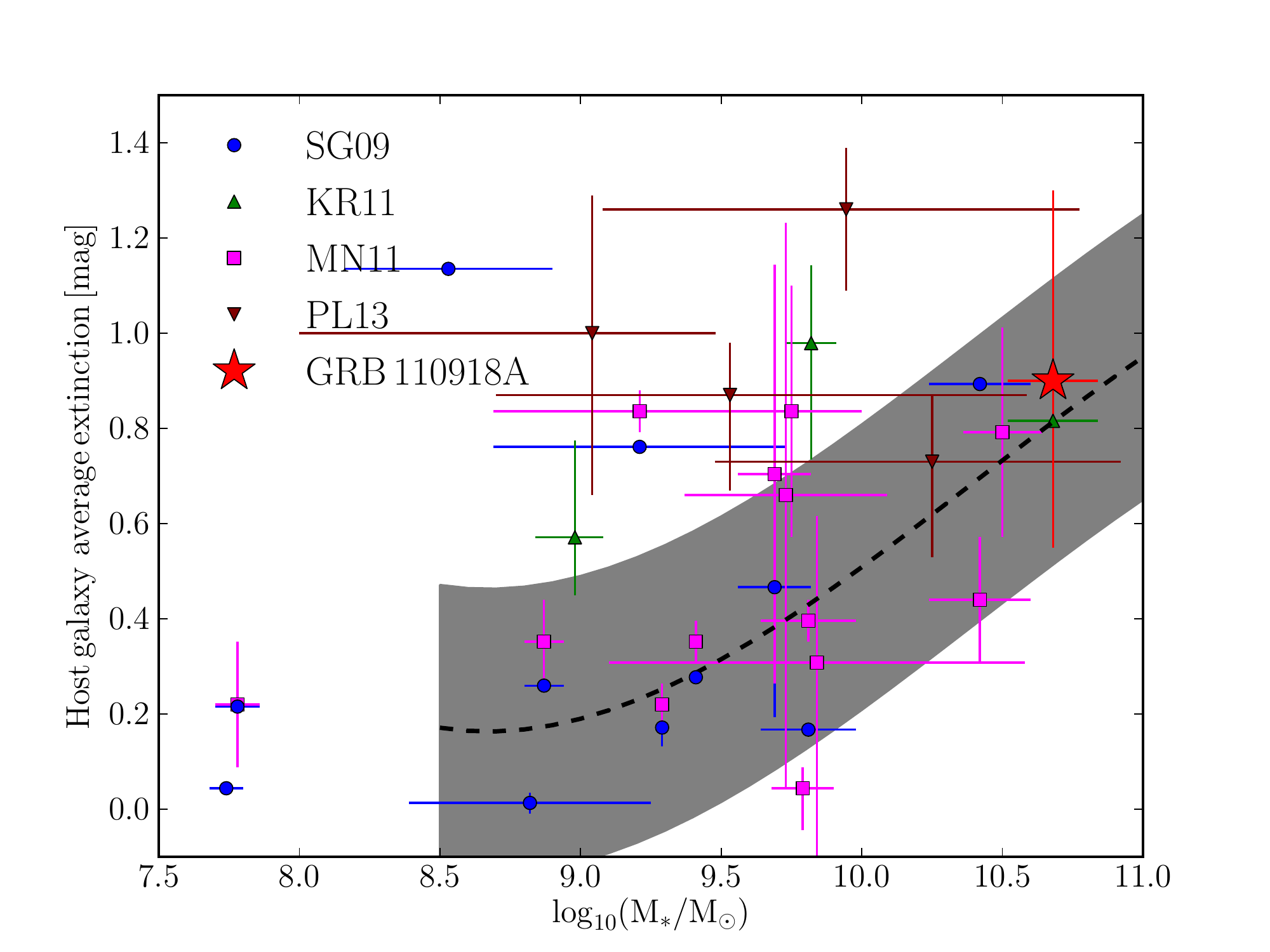}
 \caption{The average line of sight extinction of the host galaxy vs. the stellar mass. Values from SG09 and MN11 have been converted from $E(B-V)^{\mathrm{gas}}$ to $E(B-V)^{\mathrm{stars}}$ using the relation from~\citet{Calzetti00a}. The dotted-black line is the polynomial fit determined by~\citet{Garn10a} and the grey region denotes the uncertainty of $0.3\,\mathrm{dex}$.}
 \label{fig:hostav}
\end{figure}

The mass-metallicity relation of field galaxies \citep[e.g.,][]{Tremonti04a} has been studied in depth to high redshift~\citep{Savaglio05a,Erb06a,Yabe12a}. Similarly, the dust content of a given galaxy is also well known to correlate with stellar mass~\citep[e.g.,][]{Garn10a,Zahid13a}. To illustrate the behaviour for GRB hosts, we show the average host galaxy extinction versus the stellar mass of the host galaxy alongside the correlation determined by~\citet{Garn10a} in Fig.~\ref{fig:hostav}. The GRB hosts are taken from~\citet[][SG09]{Savaglio09a},~\citet[][MN11]{Mannucci11a},~\citet[][KR11]{Kruehler11a} and~\citet[][PL13]{Perley13a} and converted to a Chabrier IMF if need be. The correlation of~\citet{Garn10a} has been determined from SDSS galaxies with $z<0.7$ and so we limited our GRB sample to galaxies with $z<1.0$.
GRB hosts follow the distribution obtained from field galaxies well, with a possible excess of dusty systems at stellar masses of $10^{9-10}\,M_{\sun}$. Given the inherent systematic difficulties of determining the dust reddening in galaxies, and the heterogeneous selection of targets (in particular the KR11 and PL13 samples were initially selected to contain a lot of dust), this trend should not be over-interpreted.  What seems clear is that the host of GRB~110918A is at the high end of the distribution of stellar masses for GRB hosts, and there is no strong discrepancy between GRB-selected galaxies and field galaxies in the relation between their dust content and stellar mass.

\begin{figure}
 \includegraphics[trim=0cm 0cm 0cm 0cm, clip=True, width=9.5cm]{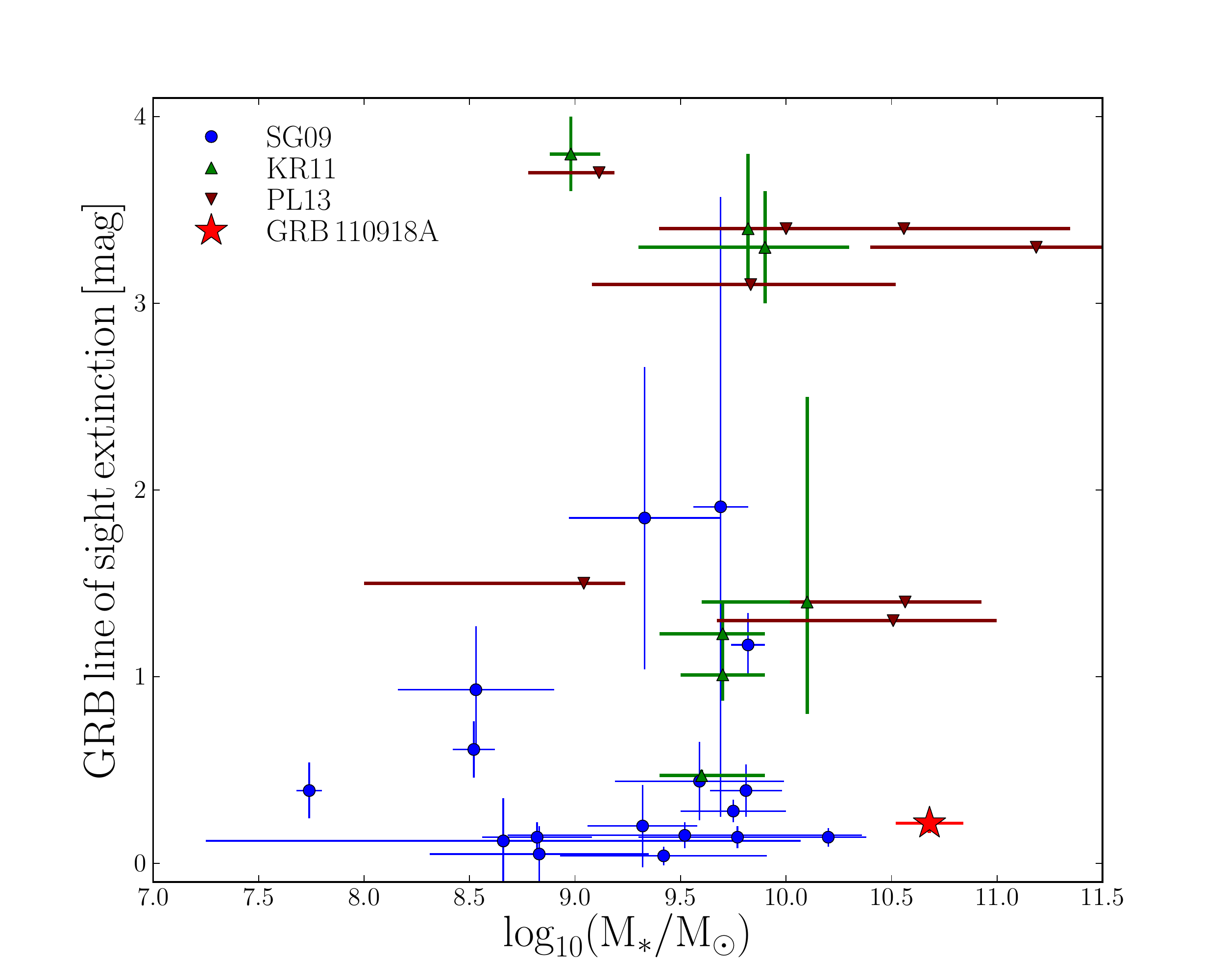}
 \caption{The GRB line of sight $A^{\mathrm{GRB}}_{V}$ plotted against the stellar mass of the host galaxy. The extinction values have been obtained from~\citet{Kann06a},~\citet{Kann10a},~\cite{Greiner11a},~\citet{Schady12a} and~\citet{Perley13a}.}
 \label{fig:grbav}
\end{figure}

Secondly we plot the host's stellar mass vs. the GRB's line of sight extinction in Fig.~\ref{fig:grbav}. \citet[][]{Perley13a} have highlighted that throughout the covered galaxy-mass scale, there is a very tight correlation between stellar-mass and sight-line extinction probed by the GRBs. Quite surprisingly, this correlation between afterglow dust and galaxy mass is found to be stronger than for any other physical property of the galaxy (PL13). From Fig.~\ref{fig:grbav} it can be seen that hosts selected due to high afterglow extinction (green, KR11; brown PL13) have systematically more massive and dust extinguished sight lines than the optically selected hosts (blue, SG09). Outliers to this trend such as GRB 061222A or GRB 100621A have already been noted~\citep[e.g.,][]{Kruehler11a,Perley09a}, that were within blue, low-mass galaxies that were locally strongly extinguished along the line of sight. GRB 110918A is the first example of a dust-poor line of sight with a galaxy mass at the high-end of the distribution (i.e., $\log_{10}\left(\frac{M_{*}}{\rm M_{\sun}}\right)>10.5$). While in principle, cases like GRB 110918A would be easy to identify (bright afterglow, easy localization, bright host), no comparable example has been reported in the literature to date.

The host of GRB 110918A shows similar host-integrated extinction ($A^{\mathrm{stars}}_{V}=0.90\,\mathrm{mag}$) to galaxies of a similar mass range (e.g., $M_{*}>10^{10}\,\mathrm{M_{\sun}}$ in Fig.~\ref{fig:hostav}  and $M_{*}>4\times10^{9}\,\mathrm{M_{\sun}}$ in Fig.~15 of~\citet{Perley13a} have an $A^{\mathrm{stars}}_{V}\gtrsim1.0\,\mathrm{mag}$). However, in comparison to the systems of similar mass, GRB 110918A exhibits at least 10 times less extinction along the GRB line of sight. Therefore, it is possible that: (i) the geometry of dust within the host of GRB110918A is more patchy than homogeneous in comparison to the rest of the massive GRB host population, in agreement with the example of GRB 100621A and 061222A, whereby clumpy dust was one explanation for having a highly extinguished afterglow within an unobscured galaxy~\citep{Kruehler11a,Perley13a}, or (ii) the progenitor had enough time to destroy local dust from its UV emission~\citep[see][and referencest therein]{Perley13a}.

\subsection{Fundamental Metallicity Relation}
\label{sec:Discussion:subsec:FMR}

The difference between galaxies of long GRBs and that of normal star forming field galaxies is still an on going debate. We have derived estimates for the mass, metallicity and SFR of the host of GRB 110918A, which facilitates comparing this galaxy with respect to normal star forming galaxies through the fundamental metallicity relation~\citep[FMR;][]{Mannucci11a}. The plane of the FMR was derived from star forming SDSS galaxies in the mass range $9.2<\log_{10}\left(\frac{M_{*}}{\mathrm{M_{\sun}}}\right)<11.4$ and is described by:
\begin{equation}
 12+\log(O/H) = 8.90 + 0.47\times(\mu_{0.32}-10),
\end{equation}
\noindent where $\mu_{0.32}=\log_{10}\left(M_{*}/\mathrm{M_{\sun}}\right)-0.32\times\log\left(\mathrm{SFR/M_{\sun}yr^{-1}}\right)$. Using the SED-determined mass and the \ion{H}{$\alpha$}-determined SFR, the metallicity from the FMR is $12+\log (O/H)=8.98\pm0.08$, in agreement with the metallicity from the \ion{N}{II}/\ion{H}{$\alpha$} line ratio of $12+\log (O/H)=8.93\pm0.13$. The method used in our metallicity estimate is the same as the one used by MN11 to construct the FMR, in order to ensure a direct comparison. The estimated errors are purely based on the uncertainties of the mass and SFR, and any systematic uncertainties from the method used to fit the stellar mass have been ignored.

\begin{figure}
  \includegraphics[width=9.5cm]{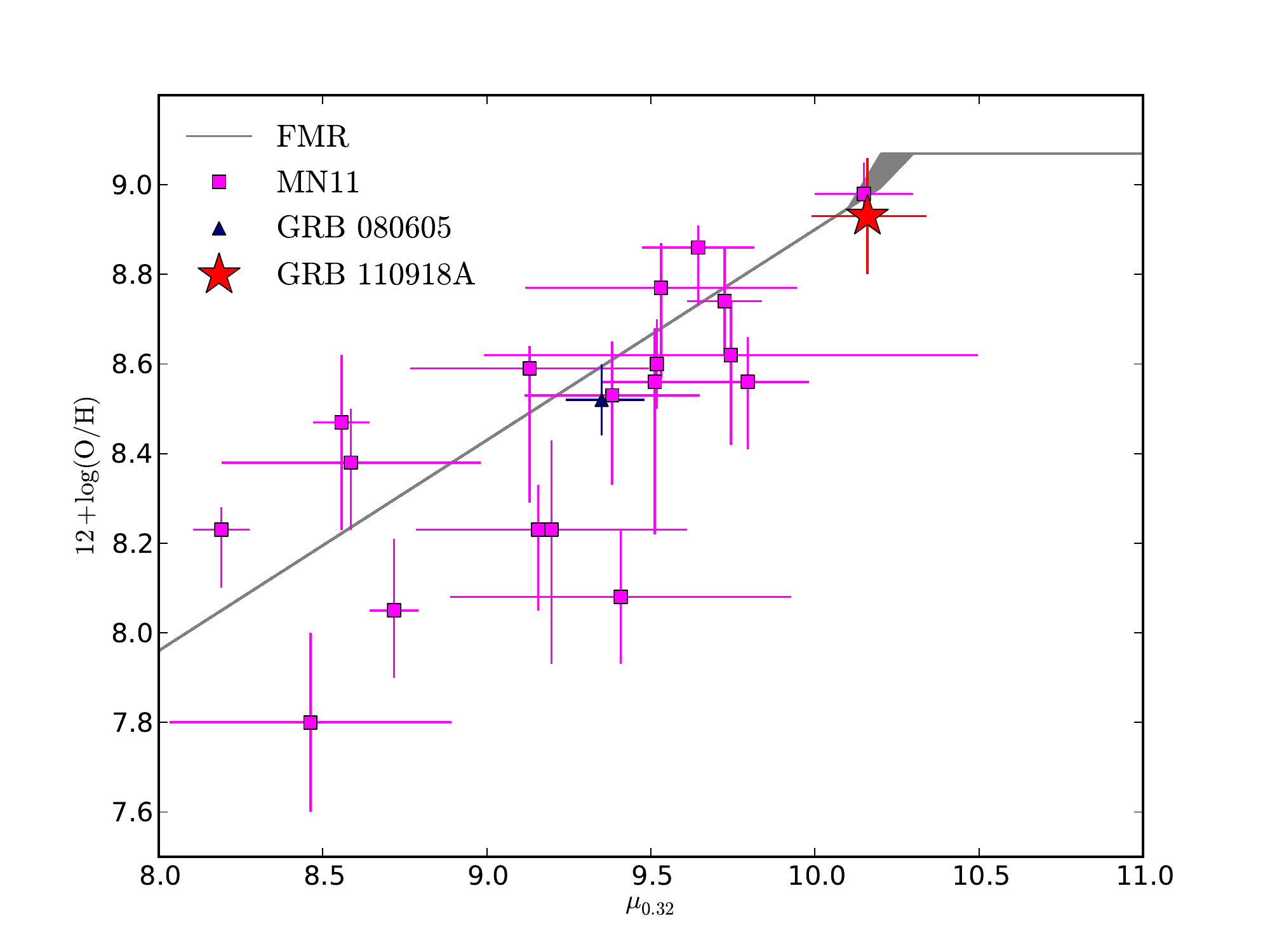}
  \caption{The metallicity determined from the fundamental metallicity relation (taken from MN11) vs. the parametric quantity $\mu_{0.32}$, plotted in grey for a range of star formation rates ($\rm SFR=0-100\,M_{\sun}yr^{-1}$). Real quantities are plotted for MN11 (magenta squares), GRB 080605~\citep[cyan upward-pointing triangle]{Kruehler12a}, and GRB 110918A (red). The host of GRB 110918A is well described by the SDSS determined FMR.}
  \label{fig:fmr}
 \end{figure}

The agreement in the characteristic properties of the host galaxy of GRB 110918A with the FMR (see Fig.~\ref{fig:fmr}) shows that the host galaxy has no deficit of metals in comparison to normal field galaxies, in line with the conclusions of~SG09, MN11, KR11, and ~\citet{Michalowski12a}. This illustrates that the mass and SFR of a GRB-selected galaxy, at least for this one event, can be used as a fair proxy for the metallicity even in the solar, or super-solar regime. 

\subsection{Metallicity and Long GRB Progenitors}
\label{sec:Discussion:subsec:metallicityandprogenitors}

\begin{figure}
 \includegraphics[width=9.5cm]{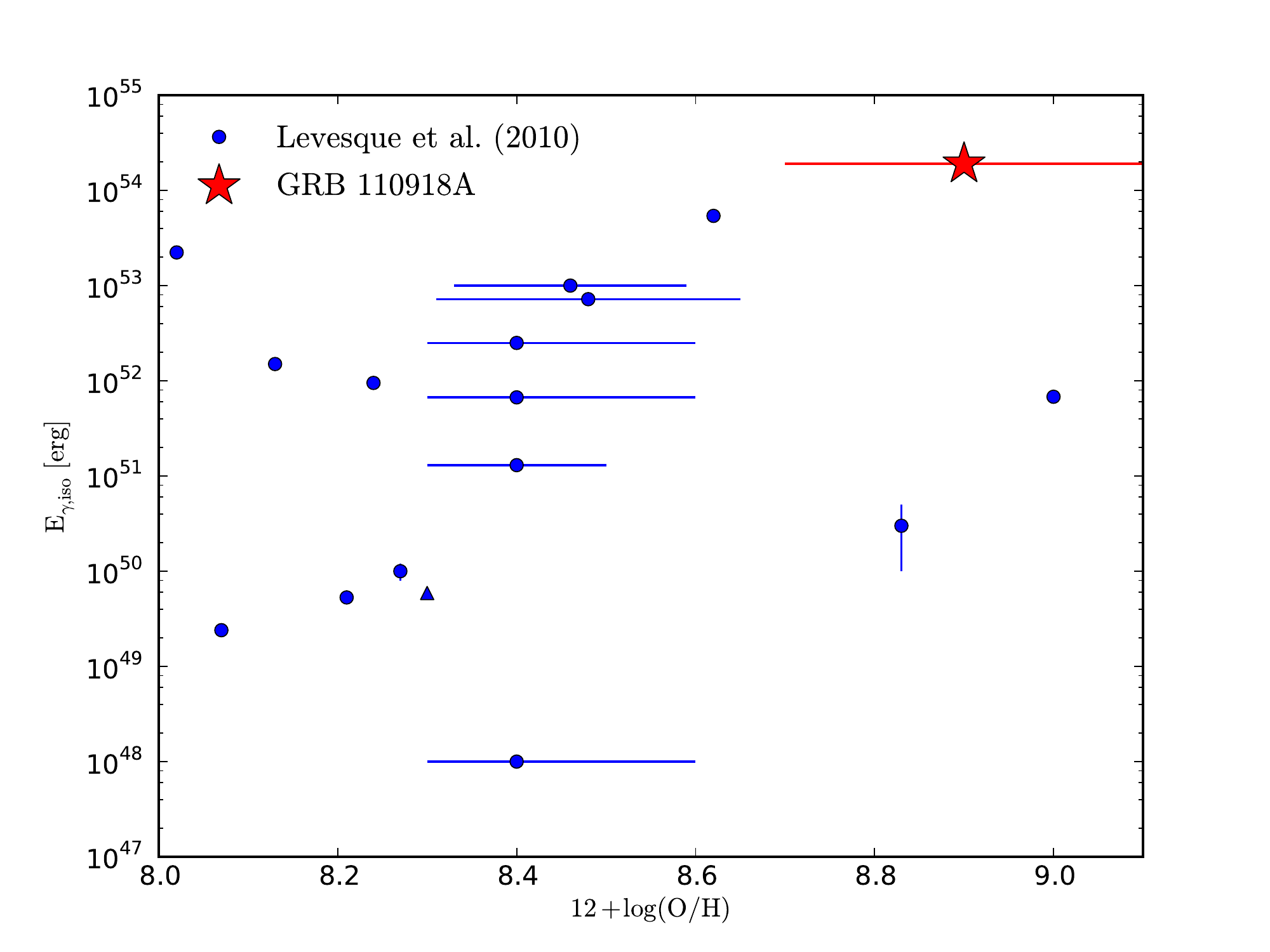}
 \caption{The isotropic-equivalent energy release in $\gamma$-rays of GRBs plotted against the gas-phase metallicity of the host galaxy. Blue data are taken from~\citet{Levesque10a}.}
 \label{fig:eiso_met}
\end{figure}

Many authors have attributed the fact that most long GRB host galaxies exhibit low metallicities as the result of an environmental preference, rather than the effect of the FMR~\citep[e.g.,][]{Modjaz08a,Graham12a,Perley13a}. This dependence on metallicity has also led to the prediction that the lower the progenitor metallicity, the larger the angular momentum, and thus the higher the energy output ($E_{\gamma,\mathrm{iso}}$) of the GRB~\citep{MacFadyen99a}. Initial studies indeed showed an anti-correlation between these two quantities~\citep{Stanek06a}, together with a cut-off metallicity above which long GRBs (for $z<0.2$) are no longer created, i.e., $Z<0.15\,\mathrm{Z_{\sun}}$. More recent studies, however, which include long GRBs at cosmological redshifts and exclude sub-luminous GRBs~\citep{Wolf07a,Levesque10a} indicate that there is no clear anti-correlation between metallicity and the GRB's energy output, as shown in Fig.~\ref{fig:eiso_met}. The prompt emission of GRB 110918A yielded an energy output of $E_{\gamma,\mathrm{iso}}=1.9\times10^{54}\,\mathrm{erg}$~\citep{Frederiks11a} within the top $2\%$ of the GRB population~\citep[][Frederiks et al. 2013]{Amati08a}. This makes GRB 110918A one of the most energetic long GRBs yet observed and its host one of the most metal rich galaxies, in contradiction to the idea of a correlation between $E_{\gamma,\mathrm{iso}}$ and metallicity.

Recently, \citet{Perley13a} performed an extensive photometric study of host galaxies selected from a sample of dark bursts, limiting the selection biases present in previous works. However, while the inclusion of dark GRB hosts increases the consistency of GRB hosts with the star formation weighted sample of field galaxies, there is still a clear lack of high-mass galaxies at $z \lesssim 1.5$. Associating the galaxy mass with metallicity, this provides indirect evidence for a metallicity effect in GRB hosts. A similar conclusion was reached based on a comparison of long GRB hosts with supernovae hosts~\citep{Graham12a}, namely that long GRB hosts show a strong preference for lower metallicity environments relative to other populations of star forming galaxies, with a metallicity cut-off of $Z<0.5\,\mathrm{Z_{\sun}}$. This cut-off is not consistent with the host galaxy of GRB 110918A, even if metallicity dispersions of $\sim0.3\,\rm dex$ are considered~\citep{Niino11a}.

\section{Conclusion}
\label{sec:Conclusion}

We observed the afterglow of GRB 110918A and its associated host galaxy and obtained photometry and spectroscopy of both. The extensive follow-up campaign has allowed us to measure the afterglow sight-line extinction as well as the attenuation of the galaxy's stellar and gas-phase component. We further derive the host's integrated SFR, stellar mass and gas-phase metallicity. In summary, this burst has revealed the following properties with respect to the long GRB population:

\begin{enumerate}

  \item The SED determined stellar mass of $\log_{10}\left(M_{*}\right/\mathrm{M_{\sun}})=10.68\pm0.16$ makes the host of GRB 110918A one of the most massive galaxies selected by a GRB at $z\sim1$.
   \item GRB 110918A is the first relatively unobscured afterglow ($A^{\mathrm{GRB}}_{V}=0.16\,\mathrm{mag}$) that has been detected in a very massive host galaxy, suggesting that the geometry of dust is more clumpy than homogeneous or local dust has been destroyed by the progenitor.
  \item The optical/NIR spectrum reveals a solar metallicity environment ($0.9-1.7\,\mathrm{Z_{\sun}}$, depending on the chosen diagnostic), making it one of the most metal-rich long GRB host galaxies found yet.
  \item Using the fundamental metallicity relation and the measured SFR, stellar mass and metallicity, we show that the host of GRB 110918A is no different to star forming galaxies selected through their own stellar light.
  \item The large energy output from the $\gamma$-ray emission of GRB 110918A and the large metallicity content of the host galaxy, is in strong contradiction with there being an anti-correlation between energy output of the GRB and environmental metallicity.
  \item Finally, the solar abundance of metals contradicts a cut-off for host galaxies of $Z<0.5\,\mathrm{Z_{\sun}}$, even if a chemical dispersion of $\sim0.3\,\mathrm{dex}$ existed.

\end{enumerate}

\begin{acknowledgements}
We thank the anonymous referees for their constructive criticisms, R{\'e}gis Lachaume for the WFI observations, and Mara Salvato and Ivan Baldry for helpful discussions. Part of the funding for GROND (both hardware as well as personnel) was generously granted from the Leibniz-Prize to Prof. G. Hasinger (DFG grant HA 1850/28-1). Based on observations made with the Gran Telescopio Canarias (GTC), installed in the Spanish Observatorio del Roque de los Muchachos of the Instituto de Astrofísica de Canarias, in the island of La Palma. This work made use of data supplied by the UK Swift Science Data Centre at the University of Leicester. This publication makes use of data products from the Wide-field Infrared Survey Explorer, which is a joint project of the University of California, Los Angeles, and the Jet Propulsion Laboratory/California Institute of Technology, funded by the National Aeronautics and Space Administration. AdUP acknowledges support from a Marie Curie Career Integration Grant Fellowship. TK acknowledges support by the European Commission under the Marie Curie Intra-European Fellowship Programme. The Dark Cosmology Centre is funded by the Danish National Research Foundation. SS acknowledges support through project M.FE.A.Ext 00003 of the MPG. FOE acknowledges funding of his Ph.D. through the \emph{Deutscher Akademischer Austausch-Dienst} (DAAD). AdUP, CT, RSR and JG acknowledge support from the Spanish research projects AYA2012-39362-C02-02, AYA2011-24780/ESP, AYA2009-14000-C03-01/ESP and AYA2010-21887-C04-01. AdUP acknowledges support from The Dark Cosmology Centre. KW acknowledges support by the STFC. PS acknowledges support by DFG grant SA 2001/1-1. DAK acknowledges support by the DFG cluster of excellence Origin and Structure of the Universe. SK, ANG, and AR acknowledge support by DFG grant Kl 766/16-1. AR, ANG, AK and DAK are grateful for travel funding support through MPE. MN acknowledges support by DFG grant SA 2001/2-1. EB acknowledges support from the National Science Foundation through Grant AST-1107973. JE is thankful for the support from A. and G. Elliott.
\end{acknowledgements}

  \bibliographystyle{aa}
  \bibliography{grb110918A}


\begin{appendix}
\label{appendix}
\section{The Afterglow Light Curve}

\begin{figure}[h!]
 \includegraphics[width=9.5cm]{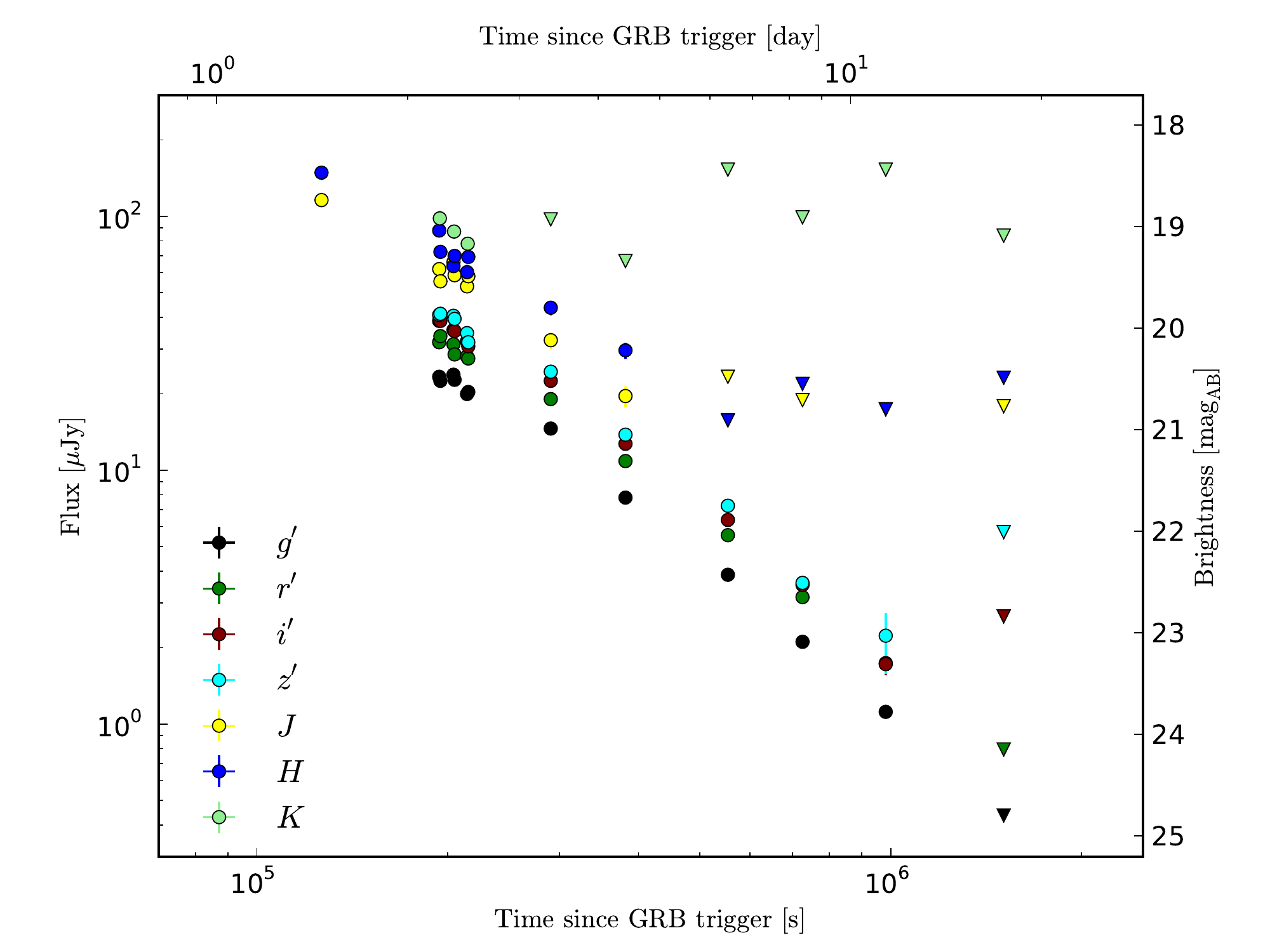}
 \caption{The afterglow light curve of GRB 110918A obtained with the 7 channel imager GROND (host subtracted).}
 \label{fig:lc_grb}
\end{figure}

The afterglow of GRB 110918A was imaged for over 40 days after the trigger with GROND in the $g'r'i'z'JHK_{S}$ bands (outlined in Sect.~\ref{sec:Observations:subsec:Afterglow:subsubsec:GROND}). Utilising the deep observations of the host, the underlying contribution from the host galaxy was subtracted using
the High Order Transform of PSF and Template Subtraction package, HOTPANTS\footnote{\url{http://www.astro.washington.edu/users/becker/hotpants.html}} \texttt{v5.1.10b}. The resulting afterglow light curve can be seen in Fig.~\ref{fig:lc_grb}, the raw data can be found in Tables~\ref{tab:stars_grboptphot} and~\ref{tab:stars_grbnirphot} and the host subtracted data can be found in Tables~\ref{tab:stars_grbsuboptphot} and~\ref{tab:stars_grbsubnirphot}. The standard stars used in $g'r'i'z'$ for relative calibration can be found in Table~\ref{tab:stars_optref}.

\begin{table*}
 \centering
 \caption{Optical reference stars.}
  \begin{tabular}{l l l l l l}
  R.A. & Dec. & $g'$ & $r'$ & $i'$ & $z'$ \\ \hline
  (J2000) & (J2000) & $\mathrm{mag_{AB}}$ & $\mathrm{mag_{AB}}$ & $\mathrm{mag_{AB}}$ & $\mathrm{mag_{AB}}$ \\ \hline 
  02:10:16.65 & -27:06:22.7 & $19.85\pm0.03$ & $19.64\pm0.03$ & $19.53\pm0.04$ & $19.54\pm0.04$ \\
  02:10:11.56 & -27:04:53.9 & $20.28\pm0.03$ & $19.86\pm0.03$ & $19.72\pm0.04$ & $19.72\pm0.05$ \\
  02:10:13.25 & -27:07:04.0 & $19.82\pm0.03$ & $19.48\pm0.03$ & $19.30\pm0.03$ & $19.27\pm0.04$ \\
  02:10:12.62 & -27:08:12.2 & $20.83\pm0.04$ & $19.37\pm0.03$ & $17.96\pm0.03$ & $17.27\pm0.03$ \\
  \hline\hline
  \end{tabular}
  \label{tab:stars_optref}
\end{table*}

\begin{table*}
 \centering
  \caption{GROND photometric data $g'r'i'z'$.}
    \begin{tabular}{l l l l l l}
      $T_{mid}-T_{0}$ & Exposure & $g'$ & $r'$ & $i'$ & $z'$ \\ \hline
      s & s & $\mathrm{mag_{AB}}$ & $\mathrm{mag_{AB}}$ & $\mathrm{mag_{AB}}$ & $\mathrm{mag_{AB}}$ \\ \hline

193892 & 691 & $20.43\pm0.04$ & $20.15\pm0.02$ & $19.92\pm0.04$ & $19.80\pm0.07$ \\
194311 & 1526 & $20.39\pm0.04$ & $20.16\pm0.03$ & $19.93\pm0.03$ & $19.73\pm0.08$ \\
194724 & 699 & $20.42\pm0.03$ & $20.18\pm0.02$ & $19.92\pm0.04$ & $19.80\pm0.06$ \\
204203 & 683 & $20.48\pm0.03$ & $20.25\pm0.02$ & $20.05\pm0.03$ & $19.87\pm0.06$ \\
204615 & 1508 & $20.48\pm0.03$ & $20.25\pm0.02$ & $20.02\pm0.03$ & $19.84\pm0.06$ \\
205025 & 689 & $20.51\pm0.03$ & $20.28\pm0.02$ & $20.02\pm0.03$ & $19.89\pm0.06$ \\
214526 & 693 & $20.56\pm0.03$ & $20.33\pm0.02$ & $20.09\pm0.04$ & $19.98\pm0.06$ \\
215017 & 1674 & $20.56\pm0.04$ & $20.33\pm0.02$ & $20.11\pm0.03$ & $19.90\pm0.07$ \\
215504 & 700 & $20.56\pm0.05$ & $20.32\pm0.03$ & $20.12\pm0.04$ & $19.91\pm0.07$ \\
290361 & 691 & $21.09\pm0.02$ & $20.81\pm0.02$ & $20.66\pm0.03$ & $20.40\pm0.06$ \\
290775 & 1519 & $21.11\pm0.03$ & $20.88\pm0.03$ & $20.65\pm0.04$ & $20.44\pm0.06$ \\
291188 & 693 & $21.09\pm0.03$ & $20.83\pm0.02$ & $20.62\pm0.03$ & $20.44\pm0.05$ \\
381456 & 1727 & $21.60\pm0.02$ & $21.30\pm0.02$ & $21.06\pm0.03$ & $20.96\pm0.05$ \\
553350 & 1732 & $22.19\pm0.04$ & $21.96\pm0.03$ & $21.59\pm0.06$ & $21.30\pm0.07$ \\
725709 & 1727 & $22.62\pm0.05$ & $22.46\pm0.05$ & $22.02\pm0.09$ & $21.54\pm0.10$ \\
981853 & 3455 & $22.89\pm0.06$ & $22.76\pm0.07$ & $22.02\pm0.09$ & $21.60\pm0.09$ \\
1507300 & 5327 & $23.09\pm0.07$ & $22.84\pm0.08$ & $22.05\pm0.07$ & $21.75\pm0.09$ \\
     \hline\hline
    \end{tabular}
  \tablefoot{
   All magnitudes have been corrected for Galactic foreground reddening. No correction has been made to subtract the flux contribution from the underlying host galaxy.
  }
  \label{tab:stars_grboptphot}
\end{table*}

\begin{table*}
 \centering
  \caption{GROND photometric data $JHK_{s}$.}
    \begin{tabular}{l l l l l}
      $T_{mid}-T_{0}$ & Exposure & $J$ & $H$ & $K_{s}$ \\ \hline
      s & s & $\mathrm{mag_{AB}}$ & $\mathrm{mag_{AB}}$ & $\mathrm{mag_{AB}}$ \\ \hline

126393 & 82 & $18.93\pm0.08$ & $18.76\pm0.08$ & .... \\
194337 & 1579 & $19.40\pm0.09$ & $19.18\pm0.12$ & $18.84\pm0.15$ \\
204643 & 1560 & $19.50\pm0.08$ & $19.19\pm0.12$ & $18.90\pm0.16$ \\
215045 & 1726 & $19.58\pm0.10$ & $19.35\pm0.12$ & $18.84\pm0.15$ \\
290803 & 1571 & $19.78\pm0.09$ & $19.64\pm0.12$ & $18.90\pm0.16$ \\
381482 & 1773 & $20.21\pm0.10$ & $19.73\pm0.16$ & $19.03\pm0.16$ \\
553375 & 1779 & $20.53\pm0.14$ & $20.17\pm0.17$ & $19.25\pm0.19$ \\
725734 & 1774 & $20.82\pm0.16$ & $20.43\pm0.21$ & $19.64\pm0.19$ \\
981879 & 3455 & $21.08\pm0.15$ & $20.59\pm0.18$ & $19.56\pm0.28$ \\
1507330 & 5379 & $20.77\pm0.12$ & $20.65\pm0.16$ & $19.50\pm0.27$ \\
     \hline\hline
    \end{tabular}
  \tablefoot{
   All magnitudes have been corrected for Galactic foreground reddening. No correction has been made to subtract the flux contribution from the underlying host galaxy.
  }
  \label{tab:stars_grbnirphot}
\end{table*}

\begin{table*}
 \centering
  \caption{GROND host subtracted photometric data $g'r'i'z'$.}
    \begin{tabular}{l l l l l l}
      $T_{mid}-T_{0}$ & Exposure & $g'$ & $r'$ & $i'$ & $z'$ \\ \hline

193892 & 691 & $20.48\pm0.04$ & $20.14\pm0.02$ & $19.93\pm0.02$ & $19.87\pm0.03$ \\
194724 & 699 & $20.52\pm0.03$ & $20.08\pm0.03$ & $19.93\pm0.03$ & $19.86\pm0.04$ \\
204203 & 683 & $20.46\pm0.04$ & $20.16\pm0.02$ & $20.02\pm0.02$ & $19.88\pm0.05$ \\
205025 & 689 & $20.51\pm0.04$ & $20.26\pm0.02$ & $20.03\pm0.03$ & $19.91\pm0.04$ \\
214527 & 693 & $20.65\pm0.03$ & $20.28\pm0.02$ & $20.13\pm0.03$ & $20.05\pm0.03$ \\
215504 & 700 & $20.63\pm0.04$ & $20.30\pm0.02$ & $20.18\pm0.03$ & $20.14\pm0.04$ \\
290776 & 1519 & $20.99\pm0.04$ & $20.70\pm0.03$ & $20.52\pm0.03$ & $20.43\pm0.04$ \\
381457 & 1727 & $21.67\pm0.03$ & $21.31\pm0.02$ & $21.14\pm0.03$ & $21.05\pm0.04$ \\
553350 & 1732 & $22.43\pm0.04$ & $22.04\pm0.03$ & $21.89\pm0.04$ & $21.75\pm0.05$ \\
725709 & 1727 & $23.09\pm0.05$ & $22.65\pm0.04$ & $22.53\pm0.06$ & $22.51\pm0.06$ \\
981853 & 3455 & $23.78\pm0.06$ & $23.30\pm0.06$ & $23.31\pm0.10$ & $23.03\pm0.28$ \\
1507300 & 5327 & $>24.80$ & $>24.15$ & $>22.84$ & $>22.01$ \\

      \hline\hline
    \end{tabular}
  \tablefoot{
   All magnitudes have been corrected for Galactic foreground reddening.
  }
  \label{tab:stars_grbsuboptphot}
\end{table*}

\begin{table*}
 \centering
  \caption{GROND host subtracted photometric data $JHK_{s}$.}
    \begin{tabular}{l l l l l}
      $T_{mid}-T_{0}$ & Exposure & $J$ & $H$ & $K_{s}$ \\ \hline
      s & s & $\mathrm{mag_{AB}}$ & $\mathrm{mag_{AB}}$ & $\mathrm{mag_{AB}}$ \\ \hline

126444 & 393 & $18.74\pm0.05$ & $18.47\pm0.07$ & .... \\
193921 & 745 & $19.42\pm0.06$ & $19.04\pm0.05$ & .... \\
194338 & 790 & $....$ & $....$ & $18.92\pm0.08$ \\
194751 & 752 & $19.54\pm0.06$ & $19.25\pm0.06$ & .... \\
204231 & 736 & $19.35\pm0.06$ & $19.39\pm0.06$ & .... \\
204643 & 780 & $....$ & $....$ & $19.05\pm0.07$ \\
205052 & 742 & $19.48\pm0.06$ & $19.29\pm0.06$ & .... \\
214555 & 746 & $19.59\pm0.06$ & $19.45\pm0.08$ & .... \\
215045 & 863 & $....$ & $....$ & $19.17\pm0.07$ \\
215531 & 753 & $19.49\pm0.06$ & $19.30\pm0.06$ & .... \\
290803 & 1571 & $20.12\pm0.08$ & $19.80\pm0.07$ & $>18.93$ \\
381482 & 1773 & $20.67\pm0.10$ & $20.22\pm0.08$ & $>19.34$ \\
553375 & 1779 & $>20.48$ & $>20.91$ & $>18.44$ \\
725734 & 1774 & $>20.71$ & $>20.55$ & $>18.91$ \\
981879 & 3455 & $>20.80$ & $>20.80$ & $>18.44$ \\
1507329 & 5379 & $>20.77$ & $>20.49$ & $>19.09$ \\

      \hline\hline
    \end{tabular}
  \tablefoot{
   All magnitudes have been corrected for Galactic foreground reddening.
  }
  \label{tab:stars_grbsubnirphot}
\end{table*}

\section{The Afterglow's Sight-Line Spectrum}

The spectra obtained with GMOS and OSIRIS (see Sect.~\ref{sec:Observations:subsec:HostGalaxy:subsubsec:Osiris}) reveal many absorption lines of gas along the line of sight toward the afterglow, specifically of the following species: \ion{Fe}{II} (2344,2374,2382,2586,2600), \ion{Mg}{II} (2803,2796), \ion{Mg}{I} (2853) and \ion{Ca}{II} (3935,3970). The equivalent widths of the metals are listed in Table~\ref{tab:spectra_equivalentwidths}.

\begin{table*}
 \center
 \caption{Equivalent widths measured for the absorption lines of the afterglow.}
  \begin{tabular}{l l l l l l l} \\ \hline
  $\lambda_{\mathrm{obs}}$ & Feature & Contaminants & $\mathrm{EW}_{\mathrm{obs}}$ & $\mathrm{EW}_{\mathrm{rest}}$ & & $z$\\ \hline
  \AA &  &  & & \\ \hline

  $4650.4$ & \ion{Fe}{II2344.2} & \ion{Fe}{$\mathrm{II^{*}}$2345.0} & $4.3\pm0.4$ & $2.2\pm0.2$ & $0.9838$ \\
  \hline
  $4713.4$ & \ion{Fe}{II2374.5}\tablefootmark{\dag} & $....$ & $....$ & $....$ & $0.9850$ \\
  $4720.5$ & $....$ & \ion{Fe}{$\mathrm{II^{*}}$2381.5}  & $8.4\pm0.6$ & $4.2\pm0.3$ & $....$ \\
  $4727.2$ & \ion{Fe}{II2382.8}\tablefootmark{\dag} & $....$ & $....$ & $....$ & $0.9389$ \\
  \hline
  $5132.7$ & \ion{Fe}{II2586.7} & \ion{Mn}{II2594.5} & $3.7\pm0.4$ & $1.9\pm0.2$ & $0.9843$ \\
  $5157.7$ & \ion{Fe}{II2600.2} & \ion{Mn}{II2696.7}, \ion{Fe}{$\mathrm{II^{*}}$2586.7} & $6.3\pm0.5$ & $3.2\pm0.2$ & $0.9836$ \\
  \hline
  $5547.3$ & \ion{Mg}{II2796.4}\tablefootmark{\dag} & $....$ & $....$ & $....$ & 0.9837 \\
  $5540.8$ & $....$ & $....$ & $11.9\pm0.4$ & $6.0\pm0.2$ & $....$ \\
  $5560.1$ & \ion{Mg}{II2803.5}\tablefootmark{\dag} & $....$ & $....$ & $....$ & 0.9833 \\
  \hline
  $5660.2$ & \ion{Mg}{I2853.0} & $....$ & $4.3\pm0.4$ & $2.2\pm0.2$ & $0.9840$ \\
  $7806.1$ & \ion{Ca}{II3934.8} & $....$ & $4.0\pm0.5$ & $2.0\pm0.2$ & $0.9839$ \\
  $7877.6$ & \ion{Ca}{II3969.6} & $....$ & $3.3\pm0.6$ & $1.7\pm0.3$ & $0.9845$ \\
   \hline\hline
  \end{tabular}
  \tablefoot{The redshift was determined for those lines with no contaminants.
             \tablefoottext{\dag}{Blended lines.}
  }
    \label{tab:spectra_equivalentwidths}
\end{table*}



\section{The Host's Emission Lines}
\begin{figure*}
  \subfloat[]{\label{sub:1}\includegraphics[width=8cm]{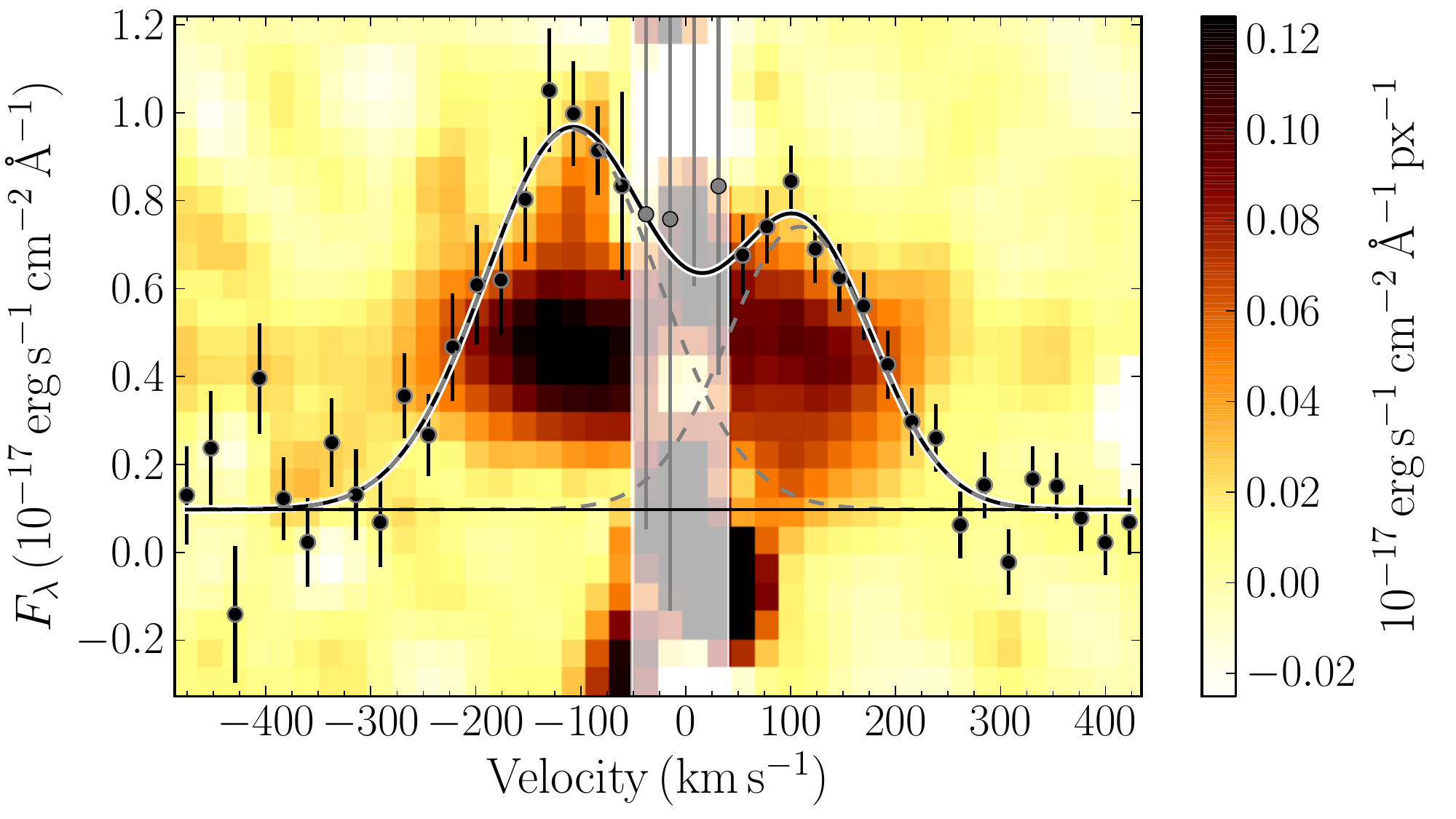}}\quad\quad
  \subfloat[]{\label{sub:2}\includegraphics[width=8cm]{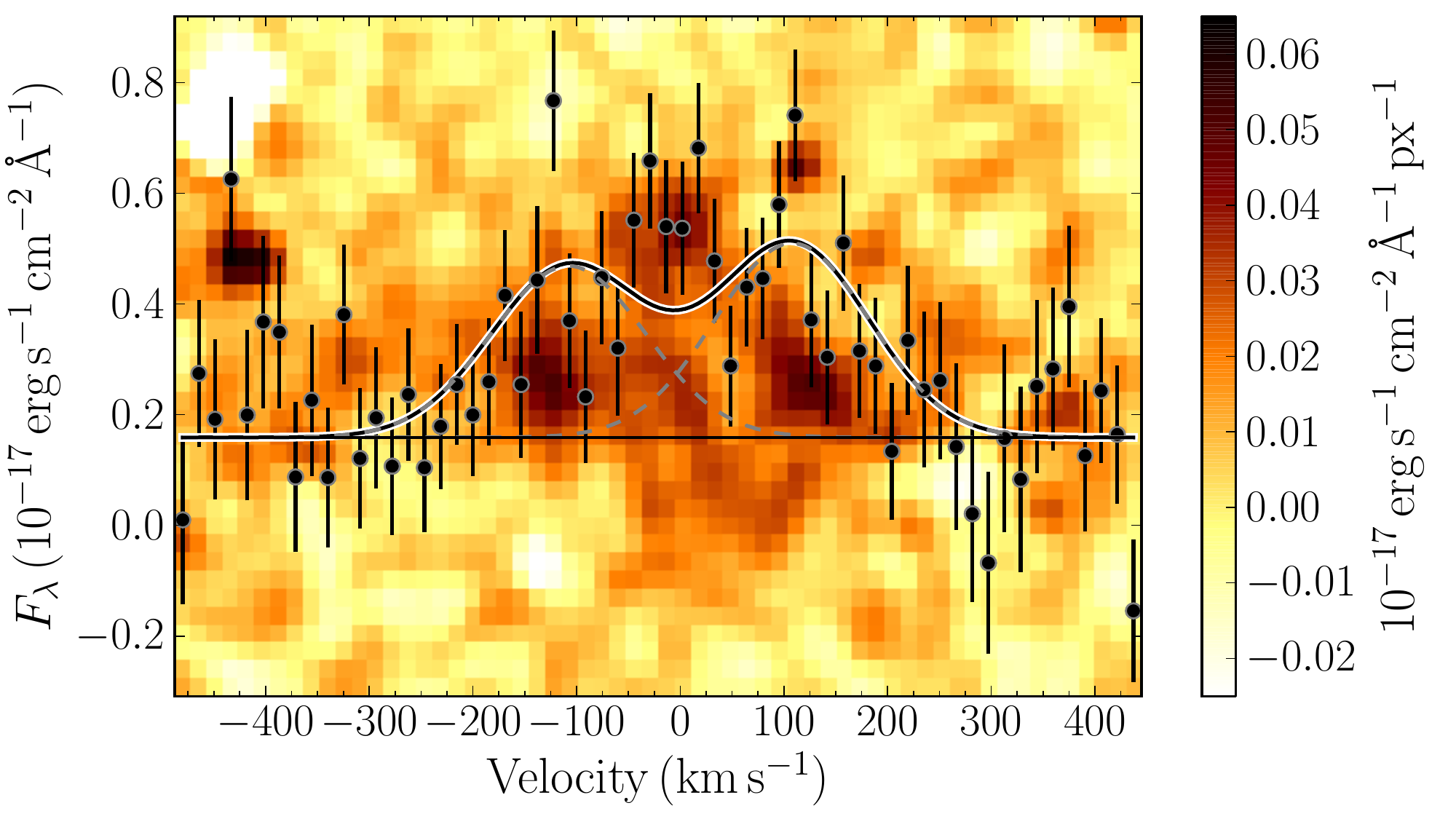}}

  \subfloat[]{\label{sub:3}\includegraphics[width=8cm]{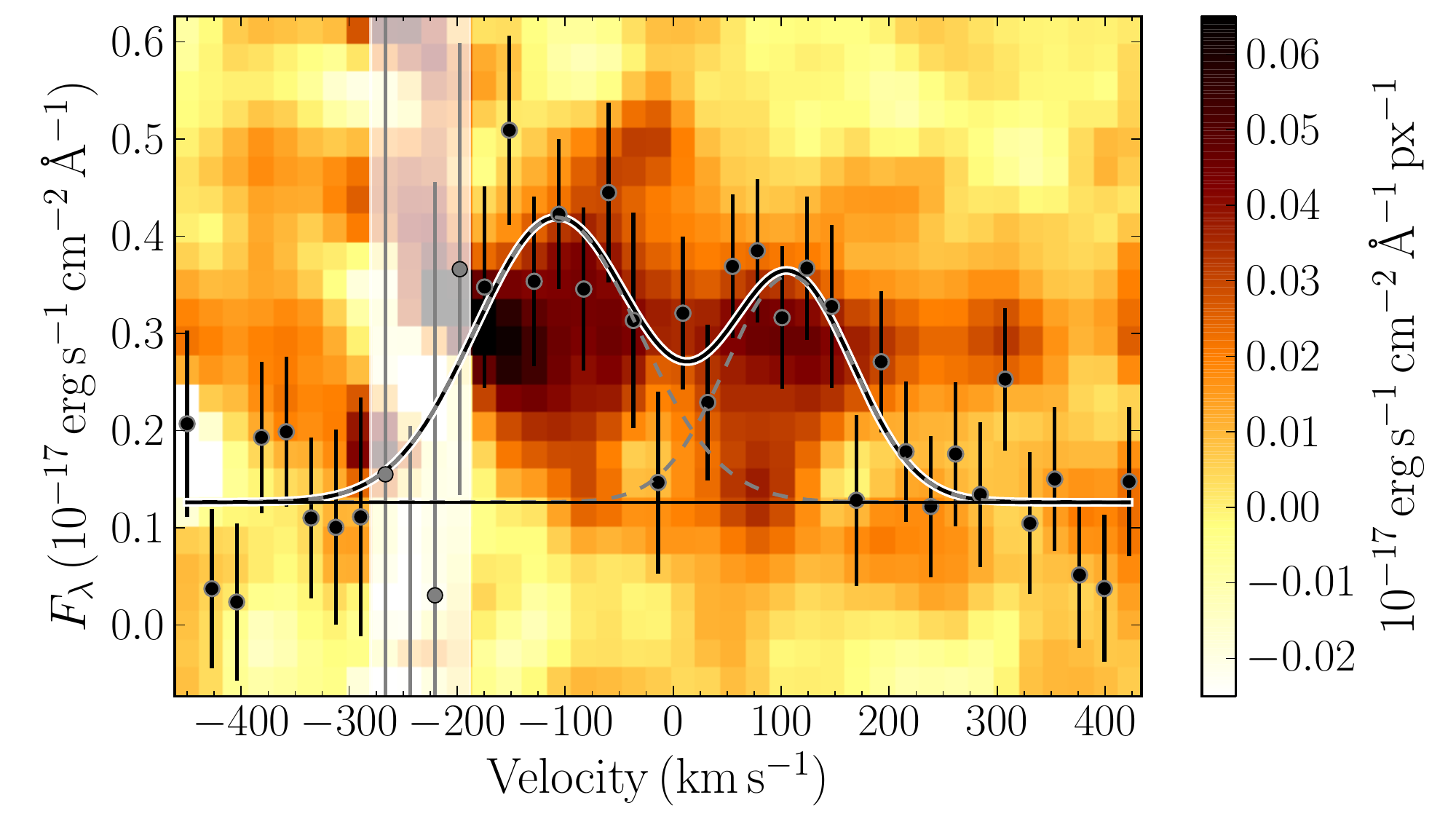}}\quad\quad
  \subfloat[]{\label{sub:4}\includegraphics[width=8cm]{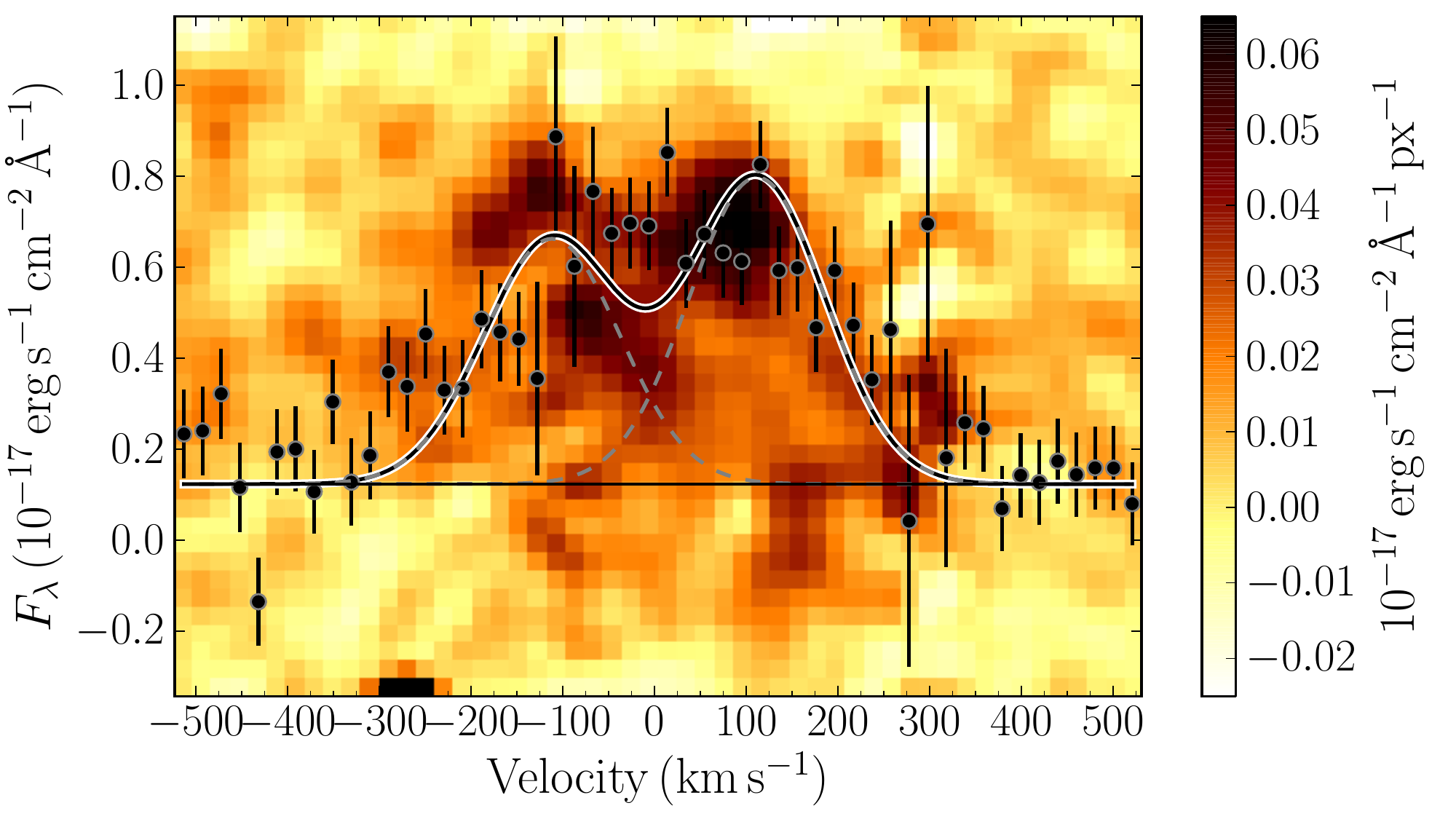}}

\caption{The 2D spectra of the host of GRB 110918A, depicting four different emissions (\nii, \oii, \ha, \hb). Overplotted is our Gaussian fit where areas that overlay telluric lines are shown in white and excluded from the fit. All the values presented are raw values and do not include slit-loss or extinction corrections. Each image has been smoothed in both pixel directions for presentation purposes. {\bf (a)}: The Balmer series transition \ha. {\bf (b)}: The Balmer series transition \hb. {\bf (c)}: The forbidden transition \nii. {\bf (d)}: The forbidden transition \oii. }
 \label{fig:lines}
\end{figure*}

Two spectra of the host galaxy were obtained with OSIRIS and X-shooter (see Sect.~\ref{sec:Results:subsec:Host:subsubsec:OpticalSpectrum}), showing the following emission lines: \ha~and \hb~transitions from the Balmer series and also forbidden transitions of \oii~and \nii~(only \oii~emission was detected with OSIRIS, and so for consistency only the X-shooter emission lines are shown). All of the 2D spectral images and 1D Gaussian fits can be seen in Fig~\ref{fig:lines}.

\end{appendix}

\end{document}